\documentclass[]{aastex62}
\usepackage{bm}

\newcommand{\funits}[1]{erg cm$^{-2}$ s$^{-1}$ \AA{}$^{-1}$}
\newcommand\xmm{\emph{XMM-Newton}}

\definecolor{bgreen}{RGB}{20,128,20}
\definecolor{purple}{RGB}{160, 32, 240}

\received{December 20, 2024}
\revised{February 25, 2025}
\accepted{March 18, 2025}
\submitjournal{The Astrophysical Journal}

\shorttitle{AU Mic \textsc{ii}: High-Frequency Radio Flares}
\shortauthors{Tristan et al.}

\begin{document}
\title{A 7-Day Multi-Wavelength Flare Campaign on AU Mic.\ II: \\
Electron Densities and Kinetic Energies from High-Frequency Radio Flares}
\correspondingauthor{Isaiah I. Tristan}
\email{Isaiah.Tristan@colorado.edu}

\author[0000-0001-5974-4758]{Isaiah I. Tristan}
\affiliation{Department of Astrophysical and Planetary Sciences, University of Colorado Boulder, CO 80305, USA}
\affiliation{Laboratory for Atmospheric and Space Physics, Boulder, CO 80303, USA}
\affiliation{National Solar Observatory, Boulder, CO 80303, USA}

\author[0000-0001-5643-8421]{Rachel A. Osten}
\affil{Space Telescope Science Institute, Baltimore, MD 21218, USA}
\affil{Center for Astrophysical Sciences, Johns Hopkins University, Baltimore, MD 21218, USA}

\author[0000-0002-0412-0849]{Yuta Notsu}
\affiliation{Department of Astrophysical and Planetary Sciences, University of Colorado Boulder, CO 80305, USA}
\affiliation{Laboratory for Atmospheric and Space Physics, Boulder, CO 80303, USA}
\affiliation{National Solar Observatory, Boulder, CO 80303, USA}

\author[0000-0001-7458-1176]{Adam F. Kowalski}
\affiliation{Department of Astrophysical and Planetary Sciences, University of Colorado Boulder, CO 80305, USA}
\affiliation{Laboratory for Atmospheric and Space Physics, Boulder, CO 80303, USA}
\affiliation{National Solar Observatory, Boulder, CO 80303, USA}

\author[0000-0003-2631-3905]{Alexander Brown}
\affiliation{Center for Astrophysics and Space Astronomy, University of Colorado, Boulder, CO 80309, USA}

\author[0000-0002-4914-6292]{Graeme L. White}
\affil{Centre for Astrophysics, University of Southern Queensland, Toowoomba 4350, Australia}

\author[0000-0001-5440-1879]{Carol A. Grady}
\affil{Eureka Scientific, Oakland, CA 94602, USA}

\author{Todd J. Henry}
\affil{RECONS Institute, Chambersburg, PA 17201, USA}

\author[0000-0002-1864-6120]{Eliot Halley Vrijmoet}
\affil{RECONS Institute, Chambersburg, PA 17201, USA}
\affil{Five College Astronomy Department, Smith College, Northampton, MA 01063, USA}

\begin{abstract}
M dwarfs are the most common type of star in the solar neighborhood, and many exhibit frequent and highly energetic flares.
To better understand these events across the electromagnetic spectrum, a campaign observed AU Mic (dM1e) over 7 days from the X-ray to radio regimes.
Here, we present high-time-resolution light curves from the Karl G.\ Jansky Very Large Array (VLA) Ku band (12--18 GHz) and the Australia Telescope Compact Array (ATCA) K band (16--25 GHz), which observe gyrosynchrotron radiation and directly probe the action of accelerated electrons within flaring loops. 
Observations reveal 16 VLA and 3 ATCA flares of varying shapes and sizes, from a short (30 sec) spiky burst to a long-duration ($\sim$5 hr) decaying exponential.
The Ku-band spectral index is found to often evolve during flares. 
{Both rising and falling spectra are observed in the Ku-band, indicating optically thick and thin flares, respectively. 
Estimations from optically thick radiation indicate higher loop-top magnetic field strengths ($\sim$1 kG) and sustained electron densities ($\sim$10$^{6}$ cm$^{-3}$) than previous observations of large M-dwarf flares.
We estimate the total kinetic energies of gyrating electrons in optically thin flares to be between 10$^{32}$ and 10$^{34}$ erg when the local magnetic field strength is between 500 and 700 G. 
These energies are able to explain the combined radiated energies from multi-wavelength observations. Overall, values are more aligned with modern radiative-hydrodynamic simulations of  M-dwarf flares, and future modeling efforts will better constrain findings.}
\end{abstract}


\section{Introduction}  
\label{sec:intro} 

{M dwarfs are the most common type of star in our galaxy \citep{Henry2024}, and stellar flares from these active objects} are thought to be one of the most efficient particle accelerators in the nearby universe \citep{Gudel1996}.
{These energetic flares} result from the explosive release of energy from the dynamic magnetic field reconnection above the stellar surface, which involves a range of physical processes including plasma heating, particle acceleration, and mass motions that produce electromagnetic radiation from the radio to X-ray regimes \citep[see][which provides an overview of our current understanding]{Kowalski2024c}.

Magnetic reconnection is thought to accelerate a beam of particles, which propagates through the plasma along a magnetic loop \citep{Fisher1985b}. While the bulk of the subsequent flare radiation is thought to be determined by the heating from the beam, only the radio emission directly probes the action of accelerated particles in stellar flares. In solar studies, hard X-ray data are {also} used to study these particle motions \citep[e.g.][]{Neidig1989, Holman2011, Kontar2011}, but detection limits {with current instrumentation} make this improbable for most stellar flares \citep{Osten2010}. Recent studies with the Atacama Large Millimeter/Sub-millimeter Array (ALMA) at 220 GHz have revealed that flaring emission is observable in the high-frequency radio and can provide constraints on magnetic field strengths and the number of relativistic electrons \citep{MacGregor2020}, but these do not give us a complete picture of the flaring evolution due to the limited frequency coverage. Previously, radio stellar flares in the optically thick regimes, which generally occur at frequencies $\lesssim 10$ GHz \citep{Nita2002}, have been used to diagnose flare properties, like radio source size and magnetic field strength, due to their expected high radiation \citep[e.g.][]{Osten2005} or detection limits \citep[e.g.][]{Leto2000}.
With new advances in radio interferometry, we are now able to reliably observe M dwarfs in higher frequencies.

Radio frequencies like the Ku (12--18 GHz) and K (18--27 GHz) bands are at a prime location to observe the optically thin regime of the radio gyrosynchrotron spectrum for the mildly relativistic electrons produced during flares \citep{Dulk1985}. {The optically thin regime lies above the peak of the spectrum, which is often taken to be $\sim$10 GHz based on solar landmark cases, despite the propensity for drastic variations in spectral peak during certain phases \citep[see Figure 1 of][]{White2003}.  Statistical studies have found that while a peak between 5 and 10 GHz is most common, peaks up to 37 GHz have been measured \citep{Nita2002, Nita2004}. As these frequencies are largely unobserved in M-dwarf flares, the extent to which this solar phenomenon is replicated in M-dwarf flares is unknown \citep{White2011, Gary2023}. Variable peaks and optically thick emissions at higher frequencies may reveal more information about these events, including source sizes and electron densities \citep[][]{Osten2005}.}

In optically thin flares, the spectral index is related to the power-law index of the electron distribution, allowing for estimation of the total kinetic energies of the gyrating electrons. Ideally, this would set the constraint on the energy partition between the accelerated particles and the multi-wavelength radiated energies (see \S{\ref{sec:disc_kinetic_energies}}) {if there are no} additions from external, thermal processes like shock heating. {This regime is thus} important for understanding the physics of flare evolution through detailed comparisons to computational models that simulate multi-wavelength flare radiation and spectra \citep[e.g.][]{Kowalski2017A, Brasseur2023, Notsu2024, Kowalski2025}.

To better understand the properties of accelerated particles in stellar flares (and how they compare to solar counterparts), we conducted a large homogeneous flare monitoring campaign of a well known flare star {including radio observations between 12 and 27 GHz}.
The first paper \citep[hereafter, T23;][]{Tristan2023} presents an analysis of the thermal empirical Neupert effect between the X-ray Multi-Mirror Mission (\xmm{}) XMM OM UVW2 (1790--2890 \AA{}) and XMM EPIC-pn X-ray (0.2--12 keV) emissions during this campaign on the dM1e star AU Mic. 
This also includes simultaneous flare frequency analysis using optical data from the Las Cumbres Observatory Global Telescope network (LCOGT).
AU Mic is a well-studied dM1e star with a radius of $0.75\pm0.03$ solar radii and two known planets \citep{Plavchan2020, Martioli2021}. It is located at a distance of $9.72$ pc \citep{Gaia, GaiaDR2}, and is a member of the
 $\beta$ Pictoris moving group with an age of $23\pm3$ Myr \citep{Mamajek2014}. AU Mic is known for its activity and strong magnetic field \citep{Klein2022}. The photospheric magnetic fields of AU Mic are split between many small-scale magnetic fields with an average value around 2600 G and a large scale component around 500 G \citep{Donati2023}.
This is the second paper in this series on our multi-wavelength campaign of AU Mic during 2018 October (Oct). Here, we continue on to analyze the simultaneous campaign radio data from the Karl G.\ Jansky Very Large Array (VLA) and Australia Telescope Compact Array (ATCA).

This paper is summarized as follows. Section 2 (\S{\ref{sec:observations}}) details the radio data reduction, including data quality assessments and considerations for times of increased noise. In \S{\ref{sec:methods}}, we focus on the radio flares, including polarization models and spectral index evolution. This also includes estimations of electron kinetic energies from optically thin flares, along with magnetic field and source size estimations from optically thick flares. In \S{\ref{sec:discussion}}, we discuss clues about flare structure that can be gleaned from polarization estimates. Further, we discuss how the electron kinetic energies can explain the optical/NUV energies from T23 and generally agree with current M-dwarf flare models, though some caveats and improvements are warranted.

\section{Data Reduction}  \label{sec:observations}
\subsection{Karl G.\ Jansky Very Large Array / Ku Band / 12-18 GHz} \label{sec:vla}
{The VLA provides 20 hours of} Ku-band (12--18 GHz) observations between Oct 11--15 while in the D configuration {(maximum baseline length of 1.03 km)} under program 18B-193 (PI: A.\ Kowalski). 
There is 13.62 hr of on-source observing time, and the data are split between 48 subbands, each having 128 channels with widths of 1 MHz. The flux and phase calibrators for all frequencies are 0137+331=3C48 and J2040-2507 respectively. The average amplitude of J2040-2507 is 0.50 Jy, with all days having values around 10\% of this average. The rms during Oct 11 is 1.5 times lower than average, while the rms during Oct 14 is 1.5 times higher. 
Local temperatures and dewpoints are supplied by the VLA observation logs (see Appendix \ref{sec:appendix_vla_calibration}) for select times, and relative humidity values average around 80\% during Oct 13-15. In contrast, Oct 11 has a stable relative humidity around 40\%, with relatively clear skies.

VLA data are reduced using the \texttt{CASA} software \citep{CASA_2022}, version 6.6.4.
We apply the standard interferometric calibration steps of Radio Frequency Interference (RFI) checking, erratic data removal, and bandpass, flux, gain, and atmospheric calibrations (more details are given in Appendix \ref{sec:appendix_vla_calibration}).
The dataset is imaged to ensure that emission originates from the central star, with no significant contributions from the debris disk or contaminating sources.
Visibility data are fit to 10-second bins using \texttt{uvmodelfit} to calculate the flux density in the $RR$ and $LL$ correlations separately\footnote{The $RR$ and $LL$ notations stand for right-circularly polarized (RCP) and left-circularly polarized emission (LCP), respectively. Also note that the X-mode corresponds to emission where the {oscillating electric field of the wave and the external magnetic fields are perpendicular to each other}. In O-mode emission, this oscillating electric field is parallel to the external magnetic field. Note that O-mode emission is often linearly polarized, affecting Stokes V measurements.}.
The time resolution is chosen to match the \xmm{} light curves, and the reported times are centered on the bins.
The Stokes parameters for total intensity ($I$) and polarization ($V$) are calculated using $I=(RR+LL)/2$ and $V=(RR-LL)/2$.
The total circular polarization is then $\pi_c(\%)=V/I\times100$ \citep{Osten2004}.
All 10-second bins achieve a signal-to-noise ratio greater than 3 (SNR $>$ 3).
Light curves are also created using the smallest possible binning of 3 seconds.
This {reveals a small flare on Oct 11 which is} a singular, high-valued point in the 10-second light curve.
Otherwise, the higher-time-resolution light curve does not reveal major differences in the flares or quiescent levels and has higher average noises, with some time ranges during Oct 14 failing to achieve an SNR $>$ 3. 
Thus, the 10-second light curve is used for all analysis, except during the shortest flare.

\subsection{Australia Telescope Compact Array  / K Band / 16-25 GHz} \label{sec:atca}
The ATCA provides 50 hours of target observations, resulting 39 hours of good on-source time in K-band (16-25 GHz) between Oct 11--15 under program C3265 while in the 6A configuration {(maximum baseline length of 6 km)}. This is split between 2 subbands of central frequencies 16.7 and 21.2 GHz, each with 2,049 channels with widths of 1 MHz. The K band is chosen for the same reasons as the Ku band for the VLA, and observations between the two are staggered to cover a longer time frame with radio observations. The absolute flux calibrator is 1934-638 and the phase calibrator is 2058-297. 

Data reduction and calibration use standard multi-frequency synthesis tasks in \texttt{MIRIAD} \citep{Sault1995}. 
Several data gaps and times of increased relative humidity are caused by storms that took place near the array site, resulting in interruptions to observing due to safety hazards from lightning in the area. 
As this frequency range covers a strong water line at 22 GHz, atmospheric effects can be significant due to the presence of water vapor and liquid water \citep{Resch1983}.
We find the average amplitudes of 2058-297 on Oct 11, 13, 14 and 15 to be within 1\% of 0.68 Jy, while on Oct 12 the average amplitude is 14\% higher, with an rms five times larger than on other days. Observations on this day are almost entirely characterized by high relative humidity (greater than 90\%), which explains the increased scatter seen in the light curves of AU Mic for these times.

The ATCA visibility data are also fit using \texttt{uvmodelfit} with \texttt{CASA} in 1-minute time bins. {Here, the Stokes parameters are calculated as $I=(XX+YY)/2$ and $V=(XY-YX)/2$ due to ATCA's linear feed data.}
The noise reported by \texttt{uvmodelfit} is extremely small, but with a high $\chi^2$ value, indicating a possible discontinuity between \texttt{MIRIAD} and \texttt{CASA} routines.
We follow the recommendation from \texttt{CASA} to multiply the errors by $\sqrt{\chi^2}$ to account for this.
We manually calculate the rms values for the Oct 11 light curve using \texttt{MIRIAD}, and values between this and the corrected errors from \texttt{CASA} are similar within 3\%.
{The 1-minute light curves have many times that do not achieve an SNR $>$ 3. Median SNR values are 4.7 in the 16.7 GHz subband and 2.5 in the 21.2 GHz subband, compared to 14.0 in the VLA wideband. 
Less than 1\% of ATCA Stokes V values achieve an SNR $>$ 3, making these results unreliable. Ten-second-binned light curves are achievable with the ATCA data, but they do not reveal any major changes and reduce the SNR. While there are multiple data points that appear similar to the small VLA flare, the 10-second light curve reveals these to be noise. Thus, the 1-minute light curve is used for all analysis.}

\section{Radio Flare Analysis}  \label{sec:methods}

\subsection{{Description of the Flare Sample}} \label{sec:flare_lc}
\begin{figure*}[!ht]
\centering
\includegraphics[width=1\linewidth]{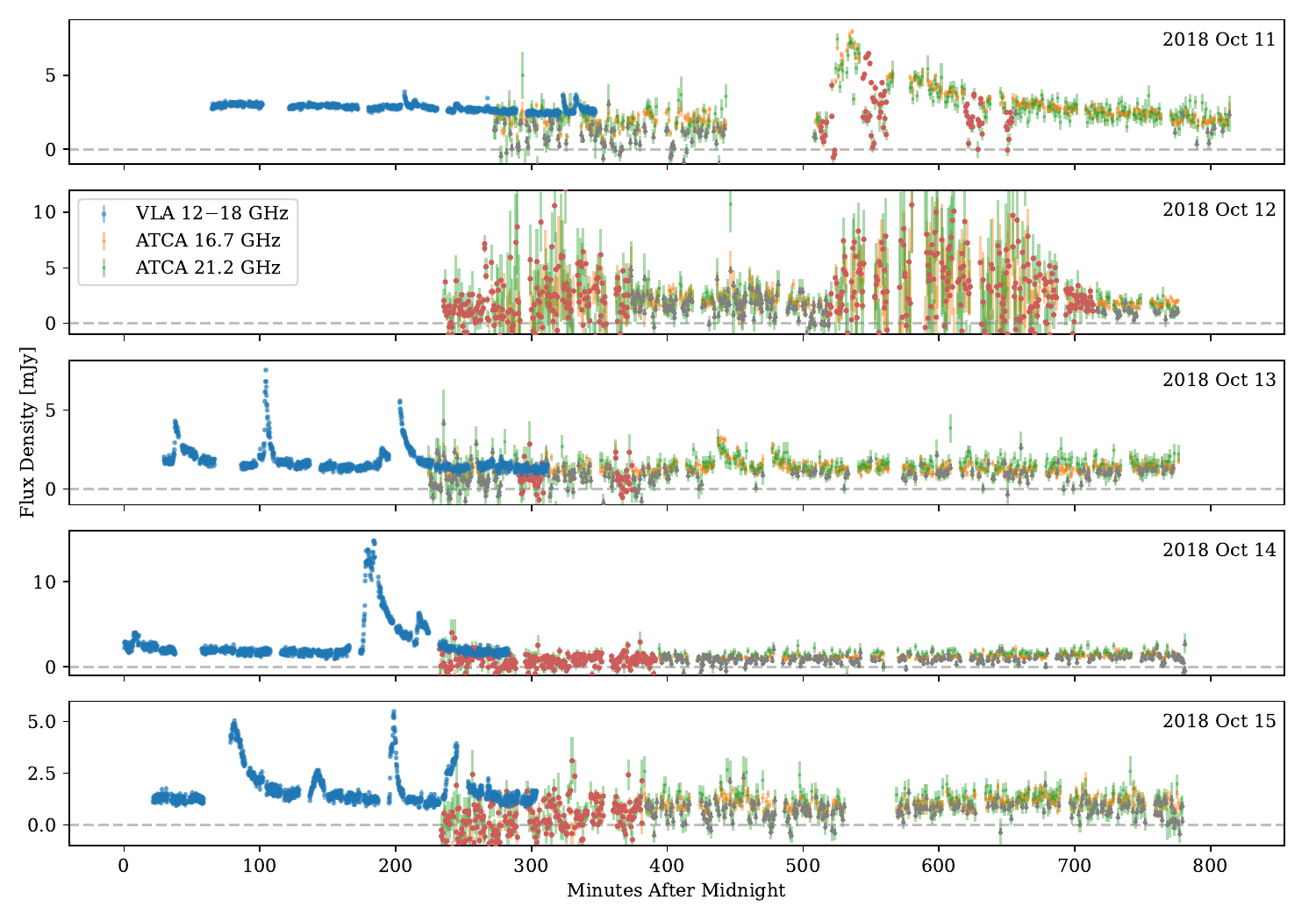}
\caption{\label{fig:RadioOnlyLCs} Light curves for the wideband VLA data and each of the two ATCA subbands. The time binning is 10 seconds for the VLA data and 1 minute for the ATCA data.
Data points that do not achieve an SNR $>$ 3 are marked with gray, downward arrows.
{ATCA intervals that are dominated by noise are marked with red circles and are not used in calculations.}
These are chosen based on the 21.2 GHz data, which are more strongly affected by the major water line in this frequency  range (see \S{\ref{sec:atca}}).
While the VLA Oct 14 data has periods of increased scatter due to weather, there are no times dominated by noise.
The gray dashed line marks a flux density of 0 mJy.
}
\end{figure*}

\begin{figure*}[!ht]
\centering
\includegraphics[width=\linewidth]{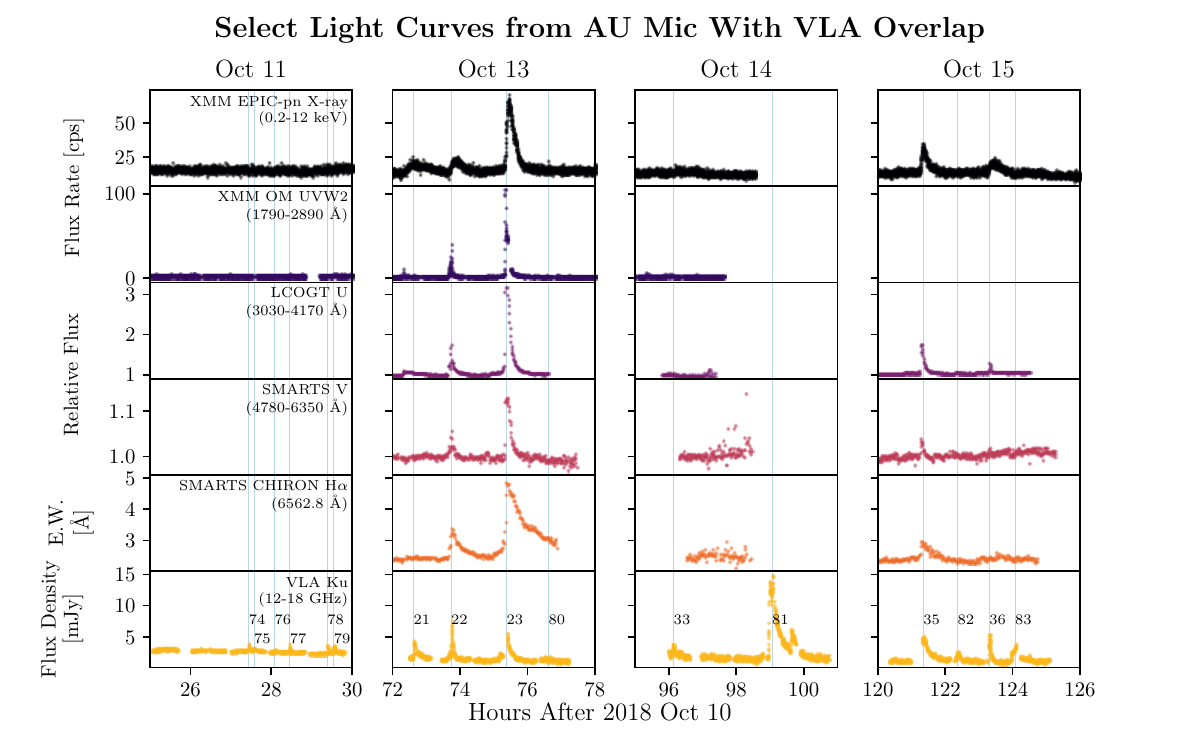}
(a)
\\[0.4cm]
\includegraphics[width=\linewidth]{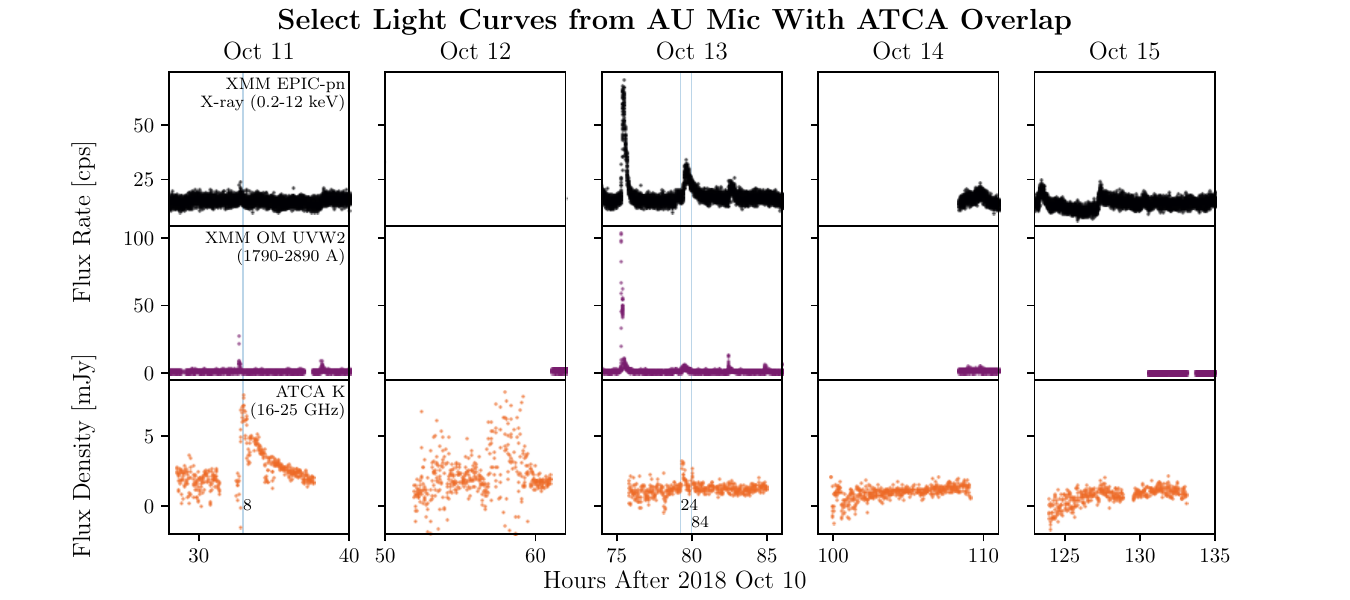} 
(b)
\\[0.3cm]
\caption{\label{fig:VLA_Multipanel_LCs}Light curves from T23 that overlap with the VLA (a) and ATCA (b) observations significantly are presented here. The XMM RGS X-ray light curve also overlaps but is not shown due to its similarity to the EPIC-pn X-ray light curve. The vertical blue lines correspond to the radio flare peaks, and Flare IDs are labeled to the right of each flare.
}
\end{figure*}

\begin{figure*}[!ht]
\centering
\includegraphics[width=0.8\linewidth]{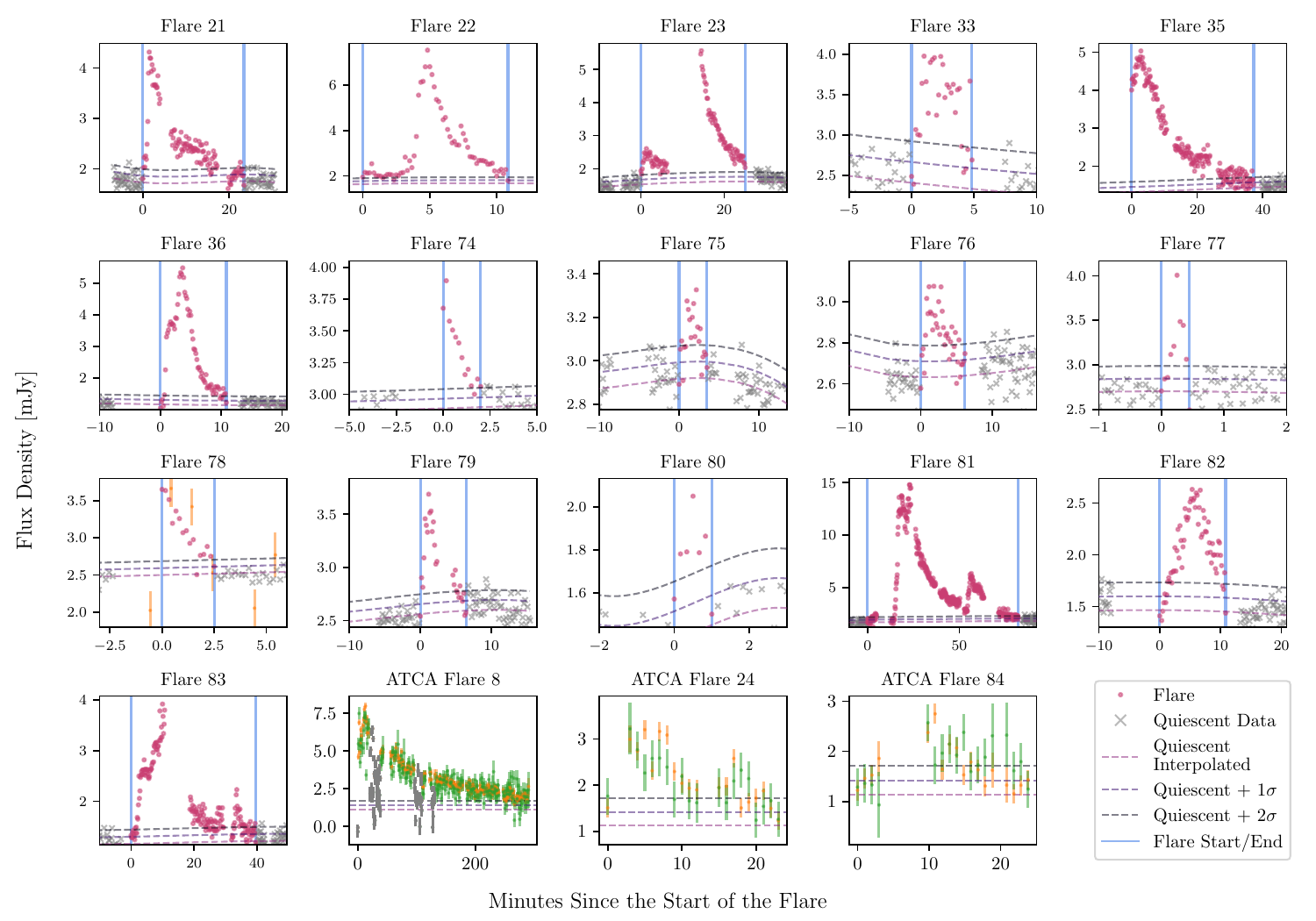}
\caption{\label{fig:VLA_Flares}
{Flares from the 10-second-binned VLA and 1-minute-binned ATCA light curves.} The ID for each flare is listed above its plot, and those below 74 indicate that the flare has multi-wavelength detections in T23. The panel for Flare 77 shows the 3-second-binned light curve. {ATCA flares show} the 16.7 GHz flux density in orange and the 21.2 GHz in green. Gray spots indicate dips due to water vapor (see \S{\ref{sec:atca}}). The ATCA 16.7 GHz data is also shown for Flare 78 to confirm that its peak lies within or very close to the VLA observation interval.
}
\end{figure*}

The 10-second VLA wideband (12--18 GHz) and 1-minute ATCA (16.7 and 21.2 GHz bands) light curves are shown in Figures \ref{fig:RadioOnlyLCs} and \ref{fig:VLA_Multipanel_LCs}, along with multi-wavelength observations from T23.
The VLA flares are selected and analyzed following \S{}2.3--4.1 of T23, and flares with IDs below 74 have previously-analyzed multi-wavelength components (cf.\ Table 6 of T23).
In summary, we select flares by analyzing time intervals with sustained flux densities above 2$\sigma$ and comparing them to multi-wavelength data when necessary. This resulted in 16 VLA flares.
The ATCA flares do not fit the previous methods well due to data gaps, noise, or a lack of surrounding quiescent areas. 
Thus, we determine ATCA flares by choosing increases in flux density where at least 2 consecutive data points rise above 2$\sigma$ from the overall quiescent flux. This method resulted in 3 clear ATCA flares. Flares 8, 24, and 84 refer to ATCA data while all other flare IDs refer to VLA flares unless stated otherwise. {Flare profiles are shown in Figure \ref{fig:VLA_Flares} and properties including timings, peak fluxes, and impulsiveness\footnote{The impulsiveness index is defined as $\mathcal{I} = I_{ \text{peak}}/t_{1/2}$, where $I(t)$ is the intensity contrast, counts$_f$/counts$_{q}-1$, $t_{1/2}$ is the full-width at half-max in minutes, $f$ corresponds to flaring emissions, and $q$ corresponds to quiescent emission \citep{Kowalski2013}.} are tabulated in Table \ref{tbl:flares}.}

A few flares warrant discussion pertaining to the data quality. Flare 8 shows multiple clear dips throughout the decay phase. These dips are deeper in the 21.2 GHz band, indicating they are likely effects from water vapor (see \S{\ref{sec:atca}}). It is questionable if this affects the flare peak, as there is a sharp drop after a single data point which rises again {only} in the 16.7 GHz observations. Regardless, the dips are removed to not skew overall statistics of this flare ({see Figure \ref{fig:RadioOnlyLCs} for more details}).
Flare 78 is the only flare with both VLA and ATCA observations, as Flare 79 has a significant ATCA data gap. {However}, there is only a marginal, 2-data-point increase in the ATCA 16.7 GHz subband. This signal implies that the VLA peak either is observed or is within 30 seconds prior to the observation. This ATCA increase is left out of further analysis, but it is shown in Figure \ref{fig:VLA_Flares} for completeness.

There are also a few marginal increases in flux density that are not reported as flares.
{One such} increase occurs on Oct 15 {and} coincides with Flare 38 in the XMM EPIC-pn X-ray data.
However, the rise and peak of this flare are lost in a data gap, there are no times when the flux density lies above 2$\sigma$, and binning the light curve to 10-seconds does not reveal any differences.  
Any {estimations made} will result in significant changes to all properties, so this response is not analyzed further. 
Likewise, there is a slight increase in flux density that correlates with XMM EPIC-pn X-ray Flare 37. 
This increase has no defined peak and is akin to other noisy data from Oct 15, so this is not considered for further analysis.
There is also a small rise at the end of the VLA Oct 15 observations indicating another flare was not observed. This flare may have coincided with a marginal increase in the XMM EPIC-pn X-ray about 15 minutes after Flare 36 that was not considered a flare. The ATCA 21.2 GHz{, but not 16.7 GHz,} data shows an additional increase 10 minutes later. Without further ways of confirming this flare, it is left out of analysis.

Flares with significant data gaps that limit their analysis due to missing rise, peaks, or decays are noted in Table \ref{tbl:flares}. 
Flares that are used for more detailed analysis in \S{\ref{sec:flare_electron_energies}--\ref{sec:flare_optically_thick}} are also marked in Table \ref{tbl:flares}.

\subsection{{Radio Cumulative Distribution}}

\begin{figure}[!ht]
\centering
\includegraphics[width=0.5\linewidth]{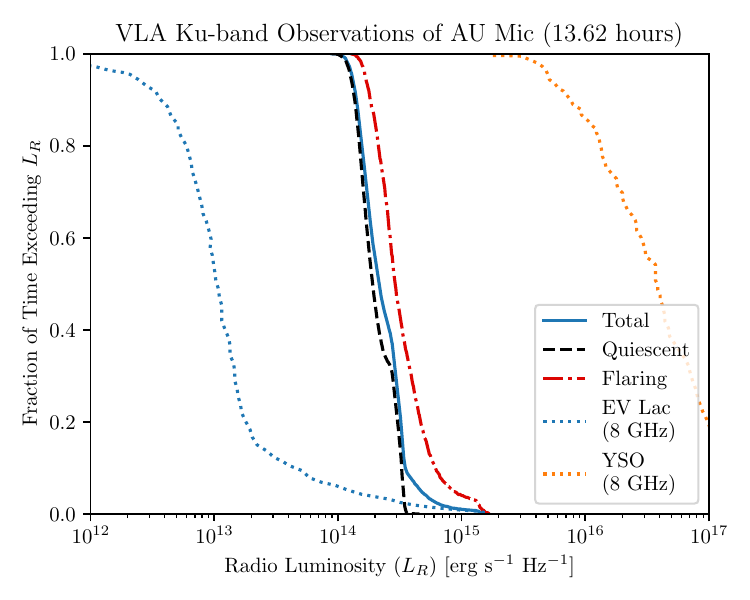}
\caption{\label{fig:luminosity_distribution}
Cumulative distributions of the VLA Ku-band (15 GHz) data, with flaring and quiescent components, compared to 8 GHz data of young solar objects (YSO) and EV Lac \citep[see][and references within]{Osten2017}. These distributions can be used to estimate the likelihood of detecting a source of $L_R$ at a certain SNR, which is useful for future mission planning, including the next-generation VLA (ngVLA). The total cumulative distribution is available online in a machine-readable format.
}
\end{figure}

A cumulative distribution function of all the data provides an overview of the variability.
Radio luminosity distributions in the form of cumulative distributions are useful for estimating the probability of observing a specific {radio luminosity ($L_R$)} and therefore detecting a desired source object within a chosen SNR \citep[][]{Osten2017}. We calculate this using the VLA Ku-band 10-second light curve of AU Mic (Figure \ref{fig:luminosity_distribution}), since all data achieve an SNR $> 3$. Separating this distribution into quiescent and flaring components reveals that the drastic change in the lower fractions is likely due to the higher quiescent level during Oct 11 (see Figure \ref{fig:RadioOnlyLCs}).

\subsection{Polarization}
\label{sec:models}

{We test multiple methods for characterizing the flaring polarization.
First, we define the fraction of total circular polarization is $\pi = V/I$, where $V = V_F + V_Q$ and $I = I_F + I_Q$. $F$ and $Q$ stand for flaring and quiescent emissions, respectively.
We create a null-hypothesis model where flaring emission is 0\% polarized ($V_F = 0$) while the quiescent emission retains its polarization level \citep[see \S{4.1.1} of][]{Osten2004}.
These models are plotted in Figure \ref{fig:FigureFlare} and the associated Figure Set.
While there are a few events that agree with the models during the rise or peak phases, these agreements are short-lived, with emission trending towards the same sense of polarization as the quiescent during the decay phase.}

We also create a grid of $\pi$ models with polarized flares, with $\pi_F = V_F / I_F$ varying between $-$100 and 100\%.
We test these models using the $\chi^2$ statistic,
\begin{equation}
    \chi^2 = \sum_{i}\frac{(\pi_{\text{data},i} - \pi_{\text{model},i})^2}{\sigma_{\pi,i}^2}, \label{eqn:chi2}
\end{equation}
with the best fitting model having $\chi^2_{\text{min}}$ and an error of values within a 90\% confidence interval for a 1-parameter model, $\chi^2_{0.9} = \chi^2_{\text{min}} + 2.71$.
Values are reported in Table \ref{tbl:flares}.
These best-fit values are consistent with simple estimations of reporting the median of $\pi_F$, which is calculated by subtracting the $I_Q$ and $V_Q$ values from $\pi$.
However, this could only be attempted for the largest flares, and the resulting errors are larger. 
These are both due to the lower SNR of the Stokes $V$ data, so the model is preferred.
Considering both methods and comparing to the unpolarized flare models, most flares do not seem to have a consistent $\pi$ value throughout, indicating varying polarization percentages with these reported values being an average. We refer the reader to \S{\ref{sec:discussion_polarization}} for an extended discussion on the polarization results.

\subsection{Spectral Index Evolution} 
\label{sec:flare_spectral_index}

The spectral index ($\alpha$) is often used as a diagnostic on whether gyrosynchrotron emission from electrons is in the optically thin or thick part of the radio spectrum and is defined as
\begin{equation}
    \alpha = \frac{\log_{10}(F_{\nu_2}/F_{\nu_1})}{\log_{10}(\nu_2/\nu_1)}, \label{eq:specind}
\end{equation}
where $F$ is flux density and $\nu$ is the representative frequency where $\nu_1 < \nu_2$.
We calculate $\alpha$ through the bandwidth of the receiver for the total flux density during each time bin. 
We use Scipy's \texttt{curve\_fit} to calculate an inverse variance weighted linear fit across 4 equally spaced frequencies.
Many flares are statistically consistent with $\alpha \simeq 0$, either due to the spectral peak being within the Ku band or because of an increase in noise due to calculating the Stokes I values with less data. 

Additionally, times during long, low-valued increases and decays may also show spectral inversions or an increase in higher frequencies. These spectral shapes may be due to the quiescent gyroresonance overpowering the flaring properties at these periods \citep[see][]{Leto2000}, but a detailed discussion of quiescent properties is beyond the scope of this study.
To account for the potential quiescent interference, we calculate the flare-only spectral index, $\alpha_F$, by subtracting the local quiescent flux density {per quadrant}. Times that return nonsensical values due to negative flux densities or extremely large uncertainties are removed.

Some, but not all, flares show statistically significant values of $\alpha$ or $\alpha_F$ that are consistent with being optically thin ($\alpha < 0$) or optically thick ($\alpha > 0$).
These can vary throughout individual flares, {though values are largely focused within the $-2.5 < \alpha_F < 2.5$ range. Evolutions of $\alpha$ per flare are shown} in Figures \ref{fig:FigureFlare}, \ref{fig:FigureFlare36}, and the associated Figure Set 1.
Of note, Flares 23 and 81 (the largest VLA flares) both show significant amounts of optically thin emission.
Overall, 4 flares have significant time intervals with $\alpha \ll 0$ {(Flares 21, 23, 35, and 81)} and 2 flares with $\alpha \gg 0$ {(Flares 22 and 36)}.
This indicates that the spectral peak can rise above, or lie below, the Ku band during M-dwarf flares.
We analyze these two sub-samples of flares further in Sections \ref{sec:flare_electron_energies} and \ref{sec:flare_optically_thick}.

\subsection{Nonthermal Energies in Optically Thin Flares} \label{sec:flare_electron_energies}

The radiated energies in the radio bands ($E_{\text{Band}}$) are calculated by multiplying the fluence (in units of erg cm$^{-2}$ Hz$^{-1}$) by $4 \pi d^2$ ($d=9.72$ pc for AU Mic) and the bandpass width, $\Delta \nu$. We use $\Delta \nu = 6$ GHz and 2 GHz for the VLA and ATCA bands, respectively. 
This calculation assumes that the emission is isotropic, though there is evidence for directivity in radio flare emission on the Sun \citep{Fleishman2003}.

We calculate a VLA flare-frequency diagram (FFD) using the on-source observing time in \S{\ref{sec:vla}}. The slope ($\beta \approx -0.7$) of the Ku-band FFD is consistent with the previous {FFDs from this campaign (see Figure \ref{fig:FFD}), confirming common average flaring rates for AU Mic from the radio to X-ray.} For comparison, we also calculate an FFD using non-simultaneous, 2-minute-cadence TESS data of AU Mic \citep[Sectors 1 and 27;][]{Ricker2015}, {and this follows a similar slope.}
The TESS FFD data reduction and caveats are discussed in Appendix \ref{sec:appendix_tess}.

The emitted energy is systematically several orders of magnitude smaller than optical or X-ray emissions. The radio data nonetheless provide important probes of a highly energetic electron population. In the chromospheric evaporation model, the total energy of accelerated particles should be comparable to the combined optical and UV radiated energy, as it represents a major source of energy for powering the flare after magnetic reconnection. This holds in some solar flare studies \citep[e.g.][]{Emslie2012}, but it is not ubiquitous \citep[e.g.\ Figure 13 of][]{Warmuth2016}. We follow the methods of \citet{Gudel2002} and \citet{Smith2005} to estimate the kinetic energies of the emitting electrons in optically thin flares using the radio light curve \citep[also see \S{3.3.5} of][replicated below]{Osten2016}.

We model the radio spectrum for gyrosynchrotron emission from mildly relativistic electrons in a homogeneous source \citep{Dulk1985} as 
\begin{eqnarray}
    S_\nu &\propto& C_1 \nu^{\alpha_{1}} \text{ for } \nu \leq \nu_{\text{peak}}\\
    S_\nu &\propto& C_2 \nu^{\alpha_{2}} \text{ for } \nu \geq \nu_{\text{peak}}\\
    \alpha_1 &=& 2.5 + 0.085 \delta_r \label{eqn:optically_thick_alpha} \\
    \alpha_2 &=& 1.22 - 0.90 \delta_r \label{eqn:optically_thin_alpha},
\end{eqnarray}
where $S_\nu$ is flux density, $\nu_{\text{peak}}$ is the peak frequency, $\alpha_1$ is the spectral index from optically thick emission, $\alpha_2$ is the spectral index from optically thin emission, $C_1$ and $C_2$ are scaling factors that depends on other observational parameters, and $\delta_r$ is the index of the distribution of the nonthermal electron population giving rise to radio flares.
We are limited to assume a constant $\nu_{\text{peak}} = 10$ GHz {despite} there being evidence for variability in $\nu_{\text{peak}}$ in some solar flares (see \S{\ref{sec:intro}}). {We can expand this model to all frequencies using the observed Ku-wideband flux density, $F(t)$, centered on the representative frequency $\nu_{\text{Ku}} = 15$ GHz,}
\begin{eqnarray}
    S(\nu,t) = F(t) \left( \frac{\nu^{\alpha_{1}}}{\nu_{\text{Ku}}^{\alpha_2}} \right) \nu_{\text{peak}}^{(\alpha_2-\alpha_1)} &\text{ for }& \nu \leq \nu_{\text{peak}} \label{eqn:S_1} \\
    S(\nu,t) = F(t) \left( \frac{\nu}{\nu_{\text{Ku}}} \right)^{\alpha_2} &\text{ for }& \nu \geq \nu_{\text{peak}}. \label{eqn:S_2}
\end{eqnarray}

From \citet{Dulk1985}, the flux density of optically thin emission can be described as 
\begin{eqnarray}
      S(\nu,t) & = & 2 \frac{k \nu^2}{c^2} \int{ T_b(\nu,t)\  d{\Omega} } \\
      & = & 2 \frac{\eta(\nu,t)V(t)}{d^2}, 
\end{eqnarray}
where $k$ is the Boltzmann's constant, c is the speed of light, $T_b$ is the brightness temperature, $d\Omega$ is the solid angle of the radio source, $\eta(\nu,t)$ is the gyrosynchrotron emissivity, $V(t)$ is the source volume, and $d$ is the stellar distance.
Using $d=9.72$ pc \citep{Gaia, GaiaDR2} for AU Mic, the radio luminosity, $L_R$ (erg s$^{-1}$ Hz$^{-1}$), can be described as
\begin{eqnarray}
    L_R &=& 4 \pi d^2 S(\nu,t) \label{eqn:L_with_S} \\
    &=& 8 \pi \eta(\nu,t) V(t). \label{eqn:L_with_eta}
\end{eqnarray}
We can then use the analytic expression of $\eta$ for X-mode emission \citep{Dulk1985},
\begin{eqnarray}
\eta(\nu,t) &=& B \left( 3.3\times10^{-24} \right) 10^{-0.52 \delta_{r}} (\sin \theta)^{-0.43+0.65\delta_{r}} \left( \frac{\nu}{\nu_{B}} \right)^{1.22-0.9\delta_{r}} N(t) \\
&=& A(\nu) N(t), \label{eqn:simple_emissivity}
\end{eqnarray}
where $B$ is the magnetic field strength local to the radio-emitting source, $\theta$ is the angle between the radio source and the line of sight (chosen to be 60$\degr$), $\nu_B$ is the electron gyrofrequency, $N(t)$ is the number density of electrons at time $t$, and $A(\nu)$ is used to contain the constant and frequency-dependent factors for simplicity.
We assume that $B$ does not change with time during the flare. 
Using Equations \ref{eqn:L_with_S}, \ref{eqn:L_with_eta}, and \ref{eqn:simple_emissivity}, rearranging, and integrating over frequencies to have a factor that depends only on time,
\begin{eqnarray}
    N(t)V(t) &=& \frac{\int{}L_R(\nu,t) d\nu}{8\pi \int{} A(\nu)d\nu} \\
    &=& \frac{d^2 \int{} S(\nu,t) d\nu}{2 \int{} A(\nu) d\nu}. \label{eqn:NT_time}
\end{eqnarray}

Within the flaring loop, {nonthermal electrons follow a power-law energy distribution,}
\begin{equation}
    n(E, t) = \frac{N(t) \ (\delta_r - 1)}{E_0} (E/E_0)^{-\delta_r},
\end{equation}
where {$N(t)$ is the number density of all electrons}, $E$ is the electron energy, $E_0$ is the cutoff energy usually assumed to be around 10 keV \citep[e.g.][]{Dulk1982, Dulk1985, White2011}. Multiplying this by a source volume $V(t)$ and integrating from $E_0$ to infinity gives a time-dependent electron kinetic energy of
\begin{equation}
    E_{\text{KE}}(t) = n_e(t) V(t) \frac{\delta_r - 1}{\delta_r - 2} E_0, \label{eqn:electron_energy_with_time}
\end{equation}
{where $n_e(t)$ is the number density of electrons with energies above the cutoff energy.}
Thus, the total electron kinetic energy over the course of a flare is given by
\begin{equation}
    E_{\text{KE}} = \frac{\delta_r - 1}{\delta_r - 2} E_0 \int{ n_e(t)V(t) dt}, \label{eqn:EKE}
\end{equation}
where $n_e(t)V(t)$ is defined by Equation \ref{eqn:NT_time} and $S(\nu,t)$ is defined by Equations \ref{eqn:S_1} and \ref{eqn:S_2}. 
A potential weakness of this model is that properties like magnetic field strength and line of sight likely change along loop length positions and evolve with time.
The observations must also be optically thin {with $2<\delta_r<7$} for these equations to apply. 

We apply this calculation to Flares 21, 23, 35, and 81 which show a value of alpha that is consistent with being optically thin.
Flare 81 is optically thin at most times, excluding the {early rise, secondary peak}, and final decay periods. We use this flare along with parameter ranges of $B=1$ to 4000 G and $\delta_r=2$ to {estimate} $E_{\text{KE}}$ in these observations (Figure \ref{fig:Contour}). These estimations vary over the energy range of $10^{30}$ to $10^{42}$ erg depending on the chosen parameters. Based on the $\alpha$ values in the data and assuming that $B$ is between 100 G and 4 kG, the likeliest energies are around $10^{31}$ to $10^{34}$ erg for Flare 81. An estimate for Flare 35 returns a similar contour map with $E_{\text{KE}}$ values reduced by a factor of 10.

Given that the evolution of $\alpha$ in Flare 81 is well-measured, we calculate $E_{\text{KE}}$ with the observed $\delta_r$ values, rather than a static value, using Equation \ref{eqn:optically_thin_alpha}. We select times when $2<\delta_r<7$ and employ a range of $B$ values, which returns similar $E_{\text{KE}}$ values to the contour estimates when {$B \geq 500$} G (see Figure \ref{fig:EnergyEstimates}).
Applying this method to Flares 21, 23, and 35 results in $E_{\text{KE}}$ values about a factor of $10^2$ smaller for this $B$ range. Flare 23 was only observed during the {early rise and late decay phases}, but it has comparable energies to Flares 21 and 35 due to its larger size. Note that these values are much rougher estimates as these flares are not always optically thin. Section \S{\ref{sec:disc_kinetic_energies}} discusses the assumptions further, including the dependence of our assumed value of $E_0$.

\subsection{Electron Number Densities from Optically Thick Flares}
\label{sec:flare_optically_thick}

Flares that are optically thick at these high frequencies can also provide valuable information about the flaring loops, including the magnetic field strength and electron number density at the optically thick surface (i.e.\ $\tau=1$). In the optically thick flaring range (i.e.\ $\alpha = 2.5$ to $3.1$ {for a homogeneous source} using Equation \ref{eqn:optically_thick_alpha} for $2 \leq \delta_r \leq 7$), the brightness and effective temperature are equal, allowing for some simplifications in the analytical equations from \citet{Dulk1985} that results in 
\begin{equation}
    S_\nu = 2 k_B T_{\text{eff}} \frac{\nu^2}{c^2} \frac{\pi r_s^2}{d^2},
\end{equation}
where $c$ is the speed of light, $d$ is the distance to the source, $r_s$ is the source size {scale}, $T_{\text{eff}}$ is the effective temperature, and $k_B$ is the Boltzmann constant. {We consider Flares 22 and 36 to be optically thick, due to their high positive spectral indices during their peaks which decline yet stay positive throughout most of their durations. A previous explanation for $\alpha$ values beyond the homogeneous bounds is an arcade of loops with differing magnetic field strengths contributing emissions, creating departures from a homogeneous, optically thick source \citep{Osten2005}. However, we observe the spectral peak falling to lower frequencies, which can also indicate internal density changes within the optically thick parts of the source. This would lower $\alpha$ values without necessarily needing changes in magnetic fields.} 

We use the empirical Equations 37 and 39 of \citet{Dulk1985} to calculate {$r_s$, $n_e$, and $B(r_s)$. Specifically, $r_s$ is the source size scale at the optically thick surface, which we assume is around the center of the flaring loop apex, and $n_e$ is the total number density of nonthermal electrons above $E_0 = 10$ keV. Then, $B(r_s)$ is the minimum $B$ value within the middle of the flaring loop at $r_s$, modeled as $B(r_s) = B_0((r_s+r_0)/r_0)^{-3}$, where $r_0$ is the stellar radius.}
{An illustration of this geometry is shown in Figure \ref{fig:flares_geometry}, which also details relevant assumptions.}
The path length to the optically thick layer is $B/\nabla B = r_s/3$, which is denoted as $L$ in these equations (not to be confused with luminosity). We then choose $B_0=2600$ G, $\theta=60\degr$, and $\delta_r=2.5$, where $\delta_r$ is based on the optically thin flares in \S{\ref{sec:flare_electron_energies}}. \citet{Osten2005} notes that the $T_b=T_{\text{eff}}$ assumption implies that the observed emission comes from the layer of the flare where emission becomes optically thick, with an estimated peak frequency at the observed frequency. However, Flares 22 and 36 show peaks around 18 and 16 GHz, respectively, in the averaged spectra of their early-to-mid decays (see example in Figure \ref{fig:FigureFlare36}). Thus we take these to be the $\nu_{\text{peak}}$ values. Results are shown in Figure \ref{fig:flares_optically_thick}.

In both flares, the minimum $B$ rises from around 1300 to 2000 G and the $r_s$ falls from around $1.35\times 10^{10}$ to $0.5 \times 10^{10}$ cm, while the $n_e$ stays around $5\times10^5$ to $10^6$ cm$^{-3}$. These can be interpreted as the values at the layer where the flare becomes optically thick, meaning $B(r_s)$ increases with the decreasing $r_s$ as the optical depth drops towards the stellar surface. Further interpretation is made easier through direct comparison to the large EV Lac flare from \citet{Osten2005}. These flares are much smaller, increasing only a few times above the quiescent in contrast to the EV Lac flare which rose about 150 times the pre-flare values. The conditions needed for optical thickness are influenced by both the smaller size and higher frequencies observed. Previous explanations for greater $B$ values at higher $\nu$ include observing higher gyrolayers \citet{Osten2005}. This is unlikely to explain the high values seen here alone, as the difference between 5 GHz and 8.3 GHz observations of EV Lac resulted in $B=60$ and 110 G at flare peak, respectively. The optically thick $\nu_{\text{peak}}$ is influenced heavily by the electron energies and $B$. Since the equations depend on the assumed $E_0 = 10$ keV, higher $B$ values are necessary to explain the higher spectral peak. While the large flare from EV Lac is similar in size to the stellar radius ($r_s \approx r_0$), the comparatively lower increase from the quiescent in flares from AU Mic result in source sizes that are much smaller than the stellar radius ($r_s = 0.24r_0$). Thus, the high, sustained $n_e$, which is similar to the peak $n_e$ of the EV Lac flare in \citet{Osten2005}, is needed to explain optically thick conditions.

Combining {the source size scale, $r_s$,} with our previous estimations of electron kinetic energies, we make simple estimates of $n_e$ for the optically thin Flare 81. We estimate the flare volume as a loop with a flare footprint area, $A_{ft}$, from 0.01\% to 0.1\% of the observable surface area of AU Mic \citep[see][]{Hawley1992, Hawley2003}. 
{Following from the geometry in Figure \ref{fig:flares_geometry}, the loop-length is $r_s \pi$.
The area along the loop is then $A_{ft,r} = A_{ft} (B_0/B(z))$, where $z$ is the height above the stellar surface and $B(z)$ is obtained from the same model as before.
} We integrate the {areas along the loop} to determine the total volume, which results in a range of $V_{\text{max}}=10^{29}$ to $10^{30}$ cm$^{3}$ using $r_s=1.35 \times 10^{10}$ cm. This is $10^2$ to $10^3$ times larger than the volume of a solar flare \citep[e.g.][]{Fleishman2022}. Inputting this into Equation \ref{eqn:electron_energy_with_time} at max $E_{\text{KE}}(t)$ gives estimates of the maximum electron number density, $n_{e,\text{max}}$, which are shown in Figure \ref{fig:EnergyEstimates}. These range from $10^4$ to $10^{14}$ cm$^{-3}$ depending of the assumed magnetic field strength. In \S{\ref{sec:disc_kinetic_energies}}, we constrain the electron densities further.

\section{{Discussion}} \label{sec:discussion}

\subsection{{Interpreting Polarization in High-Frequency Radio Flares}}
\label{sec:discussion_polarization}

The polarization of flares can be difficult to interpret{, even in spatially-resolved solar flares \citep[see discussion in][]{Nindos2020}}. High amounts of polarization can be used to determine if emission originates from coherent emission (like plasma radiation or electron cyclotron maser emission) or act as an important diagnostic for modeling \citep[e.g.][]{Dulk1985, Smith2005, Osten2008}. Outside of these uses, stellar observations become more muddled. Optically thick gyrosynchrotron observations of a homogeneous thermal plasma are unpolarized. However, a low ($<$20\%) polarization is expected if the source is comprised primarily of incoherent, nonthermal emitters \citep{Dulk1985}. \citet{Dulk1985} also notes that when emissions are optically thin, the polarization generally follows the X-mode and can be large for gyrosynchrotron emission. In this data however, all optically thin flares also stay within the $<$20\% expected from nonthermal emission. While this may have a physical explanation, it may also come from mixed signals {within} the radio-emitting source. Significant changes in $B$, $\theta$, or $T_{\text{eff}}$ along the line of sight, or throughout the source, {can} result in different polarization measurements averaging to the observed values.

Despite difficulty in interpretation, the polarization data itself can give us clues about the flaring environment. {It is common for the sense of the polarization to match between flares and the stellar quiescence. The latter represents the orientation of the more active regions, and this varies with orbital period for AU Mic \citep{Bloot2024}. Flare 74 shows an opposite sense of polarization from AU Mic, evident }by a greater trending towards $\pi=0$ than if the flare was unpolarized (see Flare 74 in Figure Set 1). This may indicate this event happened in the hemisphere opposite of the other flares. This could also be explained by radiation that comes from a coronal region on far-side of AU Mic, which would result in a reversed viewing angle and lack of observed multi-wavelength responses.
Alternatively, this may indicate a more optically thick flaring loop than expected with rather homogeneous conditions throughout, as this would result in higher relative O-mode emissions that can cause a strong trend towards $\pi=0$ \citep{Dulk1985}.

For another case, {the large flare from EV Lac remains largely unpolarized during the optically thick peak and early decay \citep{Osten2005}. In contrast, }the rise phase of Flare 22 matches the model of an unpolarized flare while $\alpha$ is in the optically thick regime, but the polarization starts to trend towards  quiescent values well before the flare becomes optically thin. This {difference} may be due to the viewing angle {during Flare 22}, where one footprint or loop side dominates the line of sight, giving preferential measurements to its sense of polarization after the initial electron beam injection, even if $\pi \approx 0$ over the entire source region for an extended time. {In all cases, line-of-sight effects may be better understood in stellar flares through future modeling efforts of the originating gyrosynchrotron source \citep[e.g.\ with the recent modeling improvements of][]{Kuznetsov2021}.}

\subsection{Electron Kinetic Energies In Light of Multi-wavelength Observations}
\label{sec:disc_kinetic_energies}
In the chromospheric evaporation model, the radiation from the optical and near-UV emission is expected to come from electron energy deposition from the beam of highly energetic electrons produced during magnetic reconnection \citep[e.g.][]{Kowalski2025}. 
If both emissions are caused by the same population of electrons, the radio beam and optical/NUV response should be temporally close and follow similar light curves.
Stellar observations can only rely on temporal coincidence to associate radio and electron-beam-impact emissions, as opposed to solar flares which also have spatial information. Examples of close temporal correlations between bands within this dataset are shown in Figure \ref{fig:flares_zoomed_in}.
Given the optically thin flares with multi-wavelength emissions in the optical and XMM OM UVW2 bands, this dataset is uniquely positioned to determine whether the energies observed within the different bands are consistent with the chromospheric evaporation model.

We compare $E_{\text{UVW2}}$, $E_U$, and $E_V$ from Table 6 of T23 to $E_{\text{KE}}$ to estimate if the energy from gyrating electrons is adequate to explain the multi-wavelength energies observed (see Figure \ref{fig:EnergyEstimates}). 
Flares 21 and 35 lack V-band and UVW2-band energies, respectively. 
From Flare 23, $E_{\text{UVW2}}/E_{U} \sim 0.6$ gives a rough upper limit to $E_{\text{UVW2}}$ in Flare 35.
All flares with V-band energies exhibit a ratio of $E_V/E_U \approx 0.6$ \citep[see also][and references within]{Hawley2014}, making V-band contributions comparable to the UVW2. While Flare 21 does not show a response in the V-band, this is more likely due to detection limits when compared to the relative flux of the U-band responses.
Thus, this ratio of $E_V/E_U$ is used for Flare 21.
Flare 23 is incomplete in the radio and Flare 81 does not have multi-wavelength observations, making further analysis of these impossible. The timing of Flare 21 is also inconsistent with the chromospheric evaporation model, as the peak aligns more with the X-ray peak, 15 minutes after the UVW2- and U-band peaks, which means the electron beam generated would not be able to cause these emissions in sequence. The radio response of Flare 21 may be from a different source region of AU Mic that did not produce responses in other bands. Thus, the estimations of Flare 35 are the most reliable results and confirm $E_{\text{KE}} \geq E_{\text{UVW2}} + E_{\text{optical}}$ when {$B \sim 500$} G (see Figure \ref{fig:EnergyEstimates}){, with Flare 21 extending this range to around 700 G if a common source is assumed between its responses.}

There are a few assumptions worth discussion. The low-energy cutoff, $E_0$, is a poorly constrained quantity in solar flares \citep{Holman2011, Krucker2011}, and it is even more so in M-dwarf flares. It is standard practice to assume $E_0 = 10$ keV.  While the gyrosynchrotron spectra change minimally above the peak frequency for $E_0 \approx 10-500$ keV \citep[see Figure 1 of][for a homogeneous radio source]{White1992}, the effect on $E_{\text{KE}}$ is not readily apparent due to the interdependence of $n_e$ on $E_0$.  All other factors being equal, larger $E_0$ values result in much smaller $n_e$ values through Equation 5 of \cite{Holman2003}. For example, if $E_0$ is raised from 10 keV to 500 keV while $\delta_r$ is held constant at 3, the inferred value of $n_e$  would be $2.5 \times 10^3$ times lower. This implies also that lower magnetic field strengths would be needed to account for multi-wavelength energies due to the resulting drop in $E_{\text{KE}}$. However, as $n_e$ is estimated from the empirical equations in \citet{Dulk1985} {which assume $E_0 = 10$ keV}, we are unable to disentangle the value without more detailed modeling\footnote{\citet{Fleishman2022} examined best-fit parameter correlations in EOVSA radio data of a solar flare.
They note that while higher values of $E_0$ may indicate lower values of nonthermal $n_e$, their study found that high $n_e$ values in solar flares do not readily depend on $E_0$. 
This is because $\delta_r$ increases with $E_0$, diminishing its correlation with $n_e$.}.

Recent simulations have considered much larger values of the low-energy cutoff, $E_0 \approx 100-500$ keV, for M-dwarf flares. These models can effectively reproduce optical and NUV properties of M-dwarf flares with higher electron beam fluxes than solar flares \citep{Kowalski2017A, Kowalski2022, Kowalski2024}. Specifically, an electron beam with a low-energy cutoff of $E_0 = 500$ keV, an energy flux density of $10^{13}$ erg cm$^{-2}$ s$^{-1}$ above the cutoff, and $\delta=3$ corresponds to a beam density of $4\times10^8$ cm$^{-3}$ during coronal propagation \citep{Kowalski2025}. 
The estimated number density for the optically thin Flare 81 is tantalizingly close to this value when $B \approx 600$ G (see \S{\ref{sec:flare_optically_thick}}). 
{When using} a smaller value of $E_0=17$ keV for the same energy flux density {models}{, the beam densities are between $n_e=3.6 \times 10^{10}$ and $5.0 \times 10^{10}$ cm$^{-3}$ when varying $\delta$ between 3 and 4, respectively. Further lowering $E_0$ to 10 keV results in $n_e=7.8 \times 10^{10}$ to $1.1 \times 10^{11}$ cm$^{-3}$ for $\delta=3$ to 4.}
If {$300 \leq B \la 700$} G, the $n_e$ estimates for Flare 81 are well within the $10^8$ to {$10^{11}$ cm$^{-3}$} range. 
A more holistic modeling approach that connects the current NUV/optical wavelengths to the radio {frequencies is needed to constrain parameters further and confirm that all emissions are reproduced.}

The simulations of \citet{Kowalski2024} also found a good agreement with optical and NUV observations using $\delta = 2.5$ to $4$, which are corroborated by the values found in the optically thin parts of Flare 81 (see Figure \ref{fig:FigureFlare}). This is also similar to the hard power-law indices ($\delta_r=2.8\pm0.06$) estimated with optically thin flares from AU Mic at higher-frequency, millimeter wavelengths \citep{MacGregor2020}. The electron number densities estimates from mm observations (220 GHz) are much lower ($\leq10^{3}$ cm$^{-3}$) than those in this study. This difference may be due to the conditions in the flaring loops, given the possible differences in $n_e$ for the optically thin or thick flares. Alternatively, the differences may simply result from using the same homogeneous source estimates with radiation that probes a different level of the flaring loop.

\section{{Summary \& Conclusions}}
\label{sec:conclusions}

For the first time, we present a target study aimed at analyzing bulk flaring properties of high-frequency radio flares in the context of a large multi-wavelength campaign. These data provide some of the first spectral constraints in the optically thin part of the M-dwarf flare gyrosynchrotron spectrum. 
We determine the spectral index evolution of flares in the Ku band, an oft overlooked frequency range for probing the physics of stellar flares. We find a mixture of optically thick and thin flares with low polarizations. From these, we estimate electron kinetic energies, magnetic field strengths, source sizes, and electron number densities. 
Radio data characterizing nonthermal emissions are also needed to constrain radiative-hydrodynamic (RHD) flaring models. Current models find that high-energy flux densities and large low-energy cutoffs are necessary to explain the thermal responses (in the optical and near-ultraviolet (NUV) wavelengths) of M-dwarf flares like those from T23. {We compare radio-band observations and evaluate whether these models are consistent with power-law index and kinetic energy calculations.} Conclusions are as follows.

\begin{enumerate}
    \item We observe 16 flares using VLA Ku-band (12--18 GHz) 10-second light curves and 3 flares using  ATCA K-band (16--25 GHz) light curves.  We find both long-lasting ($>$1 hr) and short-lived ($<$1 min) flares. The VLA flare-frequency diagram (FFD) slope is similar to other multi-wavelength regimes from AU Mic, implying common flaring rates from the radio to X-ray.
    \item We find that the circularly polarized percentages ($\pi$) of VLA flares are between 10\% and 20\%, which is consistent with expectations for optically thick gyrosynchrotron radiation. This polarization in optically thin flares may be caused by a combination of differences in parameters like $B$, $n_e$, and $\theta$ along the line of sight.
    \item Spectral index ($\alpha$) analysis of VLA flares reveals that the spectral peak is often higher than expected during M-dwarf flares ($\nu_{\text{peak}} > 10$ GHz). Many flares with $\alpha \approx 0$ are not suitable for detailed analysis due to insufficient SNR or variable conditions within the flaring loop or arcade that cause $\nu_{\text{peak}}$ to fall within the Ku band. In some cases, the peak is beyond the Ku band ($>$18 GHz); similarly high spectral peaks can sometimes occur in solar flares. This surprising property does not correlate with flare intensity.
    \item Four flares are optically thin, sustaining negative spectral indices ($\alpha<0$) over significant intervals. We estimate the electron kinetic energies of the gyrating electrons within the flaring loops ($E_{\text{KE}}$) using the inferred electron power-law indices ($\delta_r$) as a function of time. These indices are often within $2 < \delta_r < 2.6$, suggesting very hard distributions of nonthermal electrons in dMe flares. The $E_{\text{KE}}$ estimations agree well with multi-wavelength energies when using static magnetic field strengths of $500\leq B \leq700$ G. Thus, the energy pool from accelerated electrons is enough to supply the energies seen in the optical and UVW2 bands.
    \item Two flares are optically thick, sustaining positive spectral indices ($\alpha>0$) over significant intervals. We estimate larger magnetic field strengths ($B=1$ to 2 kG), smaller source sizes, and higher sustained nonthermal electron number densities ($n_e = 10^6$ cm$^{-3}$) than previous observations of an extremely large flare from EV Lac. We estimate flaring source volumes to be $10^2$ to $10^3$ times larger than solar values using a few geometry assumptions. 
    \item Combining the optically thick source volumes with the $E_{\text{KE}}$ analysis for optically thin flares, we obtain values within $n_e \approx 10^8$ to {$10^{11}$} cm$^{-3}$ for {$300 \leq B \leq 700$} G in optically thin flares. These $n_e$ and $B$ values agree with modern M-dwarf flare models that reproduce NUV and optical continuum radiation by employing powerful electron beams.
    \item The discrepancies between $B$ and $n_e$ values that are estimated from optically thin versus optically thick flares may be explained through differences in the electron low-energy cutoffs ($E_0$), magnetic field structures between the different models, {or the unique source parameters of the flares at the time of occurrence.}
\end{enumerate}

Despite overall agreements, the empirical equations that are commonly used for these estimations rely on solar parameter choices (e.g.\ $E_0=10$ keV) that may not hold in M-dwarf flares. Thus, more detailed modeling is needed following recent advancements to determine the line-of-sight effects and better constrain the parameter space. This modeling will need to be combined with current, published models using optical/NUV M-dwarf flares to not only better understand the accelerated particles within flares, but also to ensure the results are self-consistent with multi-wavelength observations. Finally, further analysis of the multi-wavelength observational data will be presented in a future paper, including an analysis of the nonthermal empirical Neupert effect.

\clearpage

\begin{deluxetable*}{lhhrrcccrrhhh}[!ht]
\tablewidth{0pt}
\tablecaption{Common Flare Properties \label{tbl:flares}}
\tablehead{
\colhead{ID} & \nocolhead{Inst.} & \nocolhead{Band} & \colhead{$t_{\text{Total}}$} & \colhead{$t_{\text{Decay}}$} & \colhead{$t_{1/2}$} & \colhead{$\mathcal{I}$} & \colhead{Peak Flux} &  \colhead{$E_{\text{Band}}$} & \colhead{$\pi_F$} & \nocolhead{Start Time} & \nocolhead{End Time} & \nocolhead{Peak Time}\\
 \colhead{} & \nocolhead{} & \nocolhead{} & \colhead{[min]} & \colhead{[min]} & \colhead{[min]} & \colhead{} & \colhead{[mJy]}  & \colhead{$10^{26}$[erg]} & \colhead{[\%]} & \nocolhead{} & \nocolhead{} & \nocolhead{}}
\startdata 
8$^{**}$ & ATCA & K 16.7 GHz & 289.7 & 277.7 & 74.1 & 0.08 & 6.83 & $84.82\pm3.24$ & \nodata & 2018-10-11T08:44:10.000 & 2018-10-11T13:33:50.000 & 2018-10-11T08:56:10.000 \\
21$^{\dag}$ & VLA & Ku & 23.4 & 21.9 & 4.2 & 0.36 & 2.60 & $6.65\pm0.44$ & $-20.8\pm2.6$ & 2018-10-13T00:36:05.000 & 2018-10-13T00:59:29.000 & 2018-10-13T00:37:35.000 \\
22$^{\dag}$ & VLA & Ku & 10.8 & 6.0 & 1.8 & 1.90 & 5.86 & $6.92\pm0.04$ & $-8.5\pm3.0$ & 2018-10-13T01:39:31.000 & 2018-10-13T01:50:21.000 & 2018-10-13T01:44:21.000 \\
23$^{*}$$^{\dag}$ & VLA & Ku & 25.0 & 10.5 & \nodata & \nodata & 3.97 & $7.26\pm3.35$ & $-13.5\pm2.2$  & 2018-10-13T03:08:39.000 & 2018-10-13T03:33:41.000 & 2018-10-13T03:23:11.000 \\
24$^{**}$ & ATCA & K 16.7 GHz & 23.0 & 18.0 & 17.2 & 0.11 & 2.07 & $3.74\pm0.22$ & \nodata  & 2018-10-13T07:14:20.000 & 2018-10-13T07:37:20.000 & 2018-10-13T07:19:20.000 \\
33 & VLA & Ku & 4.8 & 2.5 & 4.1 & 0.17 & 1.61 & $1.98\pm0.05$ & $-9.1\pm11.1$ & 2018-10-14T00:06:15.000 & 2018-10-14T00:11:05.000 & 2018-10-14T00:08:35.000 \\
35$^{*}$$^{\dag}$ & VLA & Ku & 37.2 & 34.4 & \nodata & \nodata & 3.71 & $20.17\pm0.47$ & $9.2\pm1.9$ & 2018-10-15T01:18:22.000 & 2018-10-15T01:55:36.000 & 2018-10-15T01:21:12.000 \\
36$^{\dag}$ & VLA & Ku & 10.8 & 7.2 & 4.3 & 0.86 & 4.31 & $7.21\pm0.05$ & $9.2\pm2.8$ & 2018-10-15T03:15:01.000 & 2018-10-15T03:25:51.000 & 2018-10-15T03:18:41.000 \\
74$^{*}$ & VLA & Ku & 2.0 & 1.8 & \nodata & \nodata & 1.01 & $0.37\pm0.01$ & $-19.1\pm16.7$ & 2018-10-11T03:26:25.000 & 2018-10-11T03:28:25.000 & 2018-10-11T03:26:35.000 \\
75 & VLA & Ku & 3.5 & 1.3 & 2.5 & 0.06 & 0.41 & $0.27\pm0.01$ & $21.7\pm25.8$ & 2018-10-11T03:32:25.000 & 2018-10-11T03:35:55.000 & 2018-10-11T03:34:35.000 \\
76 & VLA & Ku & 6.2 & 4.3 & 4.2 & 0.04 & 0.44 & $0.50\pm0.02$ & $-5.6\pm26.1$ & 2018-10-11T04:02:59.000 & 2018-10-11T04:09:09.000 & 2018-10-11T04:04:49.000 \\
77 & VLA & Ku & 0.5 & 0.2 & 0.2 & 2.68 & 1.30 & $0.09\pm0.01$ & $-11.1\pm39.5$ & 2018-10-11T04:27:22.500 & 2018-10-11T04:27:49.500 & 2018-10-11T04:27:37.500 \\
78$^{*}$ & VLA & Ku & 2.5 & 2.5 & \nodata & \nodata & 1.15 & $0.56\pm0.01$ & $-10.1\pm21.5$ & 2018-10-11T05:23:04.000 & 2018-10-11T05:25:34.000 & 2018-10-11T05:23:04.000 \\
79$^{*}$ & VLA & Ku & 6.4 & 5.3 & 3.9 & 0.11 & 1.12 & $1.39\pm0.18$ & $12.8\pm18.3$ & 2018-10-11T05:31:34.000 & 2018-10-11T05:37:59.000 & 2018-10-11T05:32:44.000 \\
80 & VLA & Ku & 1.0 & 0.5 & 0.9 & 0.51 & 0.64 & $0.17\pm0.01$ & $-14.1\pm18.3$ & 2018-10-13T04:37:14.000 & 2018-10-13T04:38:14.000 & 2018-10-13T04:37:44.000 \\
81$^{\dag}$ & VLA & Ku & 82.1 & 59.0 & 13.0 & 0.59 & 13.10 & $102.88\pm8.11$ & $-16.7\pm2.0$ & 2018-10-14T02:40:56.000 & 2018-10-14T04:02:59.000 & 2018-10-14T03:03:58.000 \\
82 & VLA & Ku & 10.8 & 5.5 & 7.8 & 0.10 & 1.17 & $2.64\pm0.04$ & $11.7\pm4.1$ & 2018-10-15T02:16:40.000 & 2018-10-15T02:27:30.000 & 2018-10-15T02:22:00.000 \\
83$^{*}$ & VLA & Ku & 39.6 & \nodata & \nodata & \nodata & 2.75 & $13.97\pm3.33$ & $21.0\pm2.8$ & 2018-10-15T03:54:40.000 & 2018-10-15T04:34:14.000 & 2018-10-15T04:04:40.000 \\
84$^{**}$ & ATCA & K 16.7 GHz & 23.8 & 13.0 & 10.6 & 0.13 & 1.62 & $1.92\pm0.33$ & \nodata & 2018-10-13T07:47:20.000 & 2018-10-13T08:11:10.000 & 2018-10-13T07:58:10.000 \\
\enddata
\tablecomments{
$t_{\text{Total}}$ is the flare duration, $t_{\text{Decay}}$ is the time between the peak and end of the flare, $t_{\text{1/2}}$ is the time interval at the FWHM of the light curve, $\mathcal{I}$ is the impulsiveness (see \S{\ref{sec:flare_lc}}), $E_{\text{Band}}$ is the energy within the specified band, $\pi_F$ is the percent of circularly polarized flaring emission from model fitting.
Flares with IDs $<74$ have previously published multi-wavelength components in T23.
A machine-readable table is available online with columns for Instrument, Band, Start Time, End Time, and Peak Time.
$^*$These flares have significant data gaps, and the reported timings, rate, and energies are minimum values. 
$^{**}$These are flares from ATCA observations. Values for the 16.7 GHz data are listed. The 21.2 GHz values are also listed in the online table. 
$^{\dag}$These flares were used for optically thin ($\alpha < 0$) or thick ($\alpha > 0$) flare analysis.
}
\end{deluxetable*}

\begin{figure*}[!ht]
\centering
\includegraphics[width=0.8\linewidth]{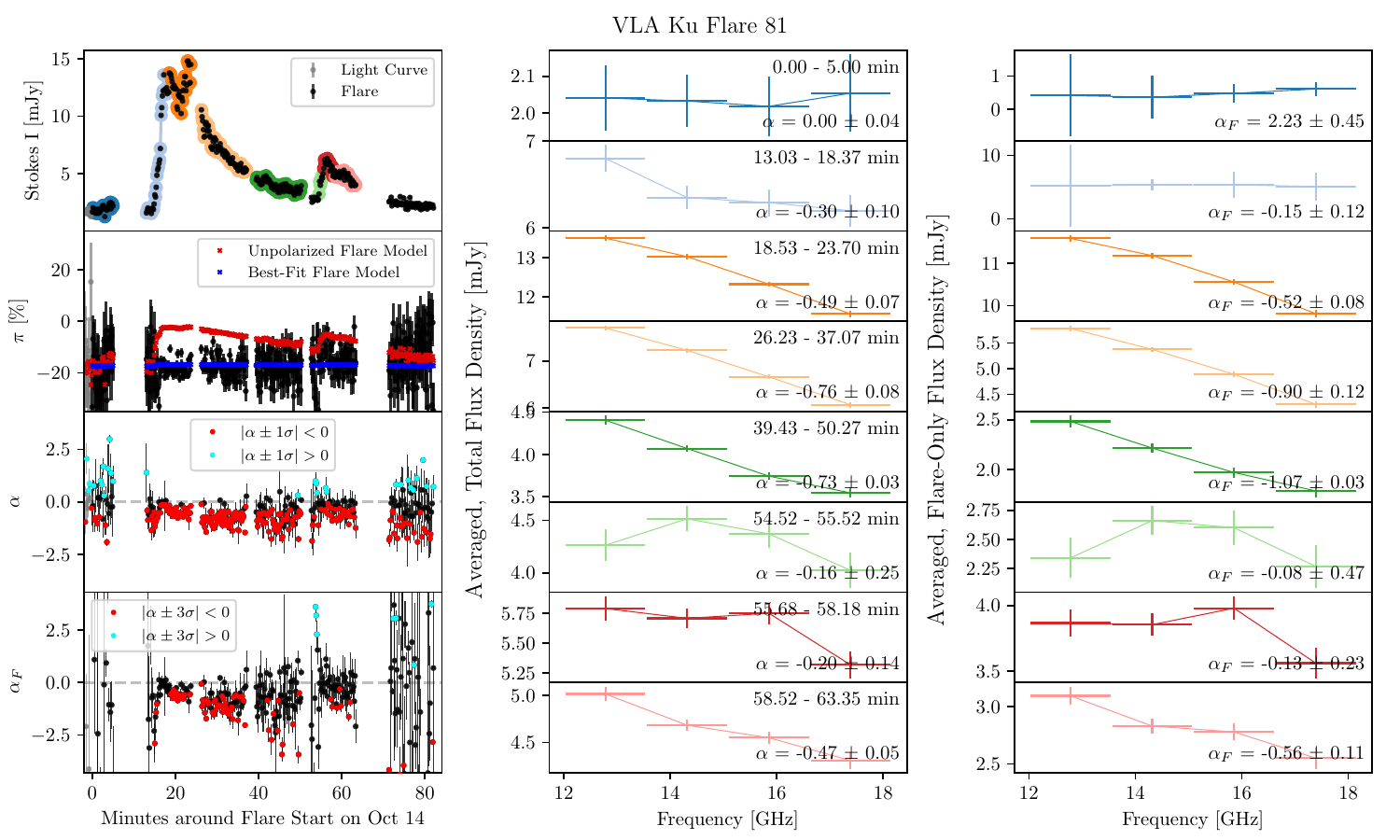}
\caption{\label{fig:FigureFlare}
The Stokes I values, percentage of circularly polarized flux ($\pi$), and spectral index of the total ($\alpha$) and flare-only flux ($\alpha_F$) are shown for the optically thin Flare 81.
For $\pi$, the red line indicates a model with $V_F=0$ while the blue line indicates the best-fit polarized-flare model with values reported in Table \ref{tbl:flares} (see \S{\ref{sec:models}}).
For $\alpha$, the bright blue and red dots indicate times when the spectral index, including errors, are in the optically thick or thin regimes, respectively. 
Note that a more stringent criterion is applied for the flare-only spectral index to emphasize the high statistical significance of this detection.
Averaged total spectra for selected intervals during the flare are shown in the middle panels, and the corresponding flare spectra are shown in the right panels.
For both, the colors correspond to the light curve regions shown in the top-left panel.
The complete figure set for all flares is available in the online article.
}
\end{figure*}

\begin{figure*}[!ht]
\centering
\includegraphics[width=0.8\linewidth]{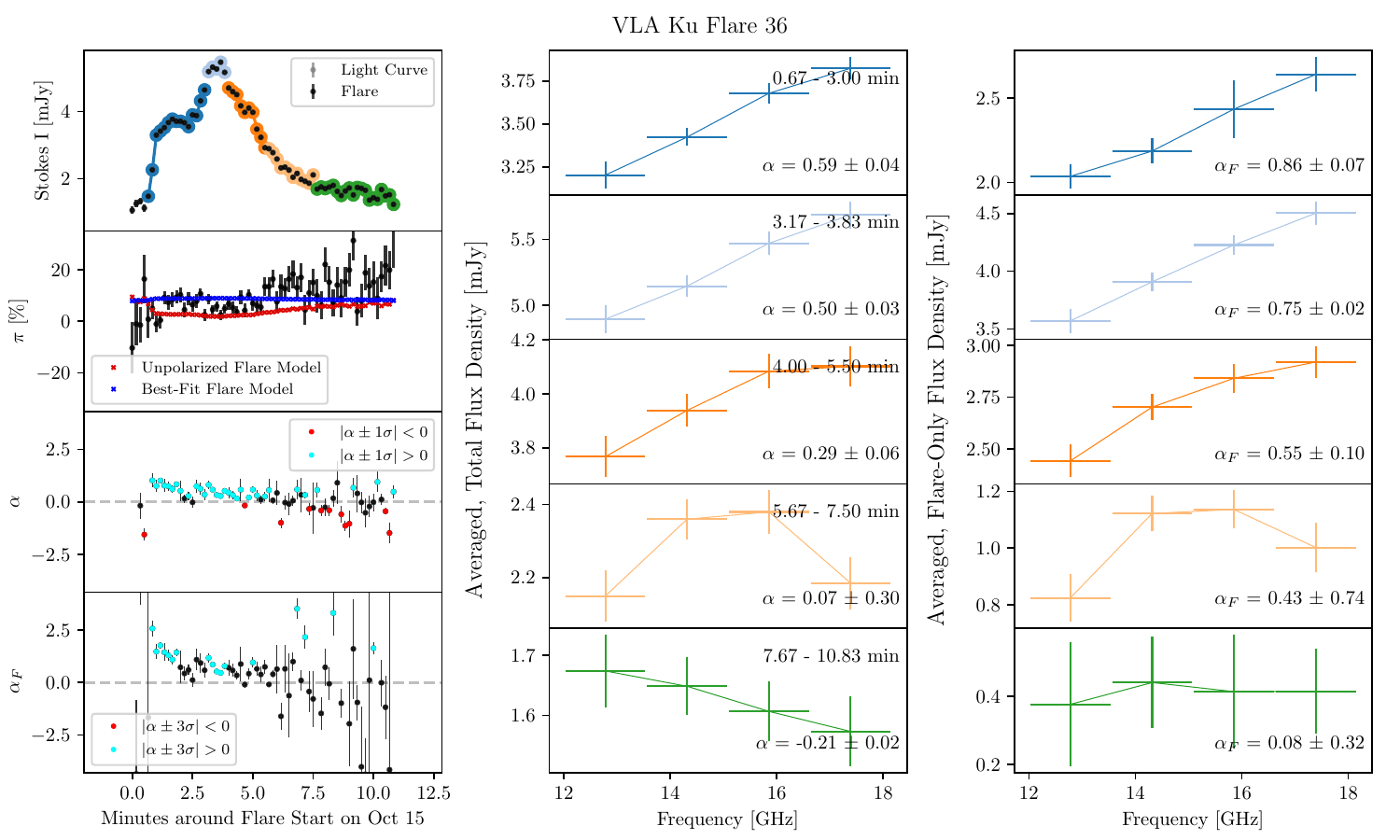}
\caption{\label{fig:FigureFlare36}
Derived Stokes I, polarization, spectral indices, and spectra for the optically thick Flare 36.
These diagrams are in the same format as Figure \ref{fig:FigureFlare} and Figure Set 1.
}
\end{figure*}

\figsetstart
\figsetnum{5}
\figsettitle{Flare Diagnostics}

\figsetgrpstart
\figsetgrpnum{5.1}
\figsetgrptitle{Flare 21}
\figsetplot{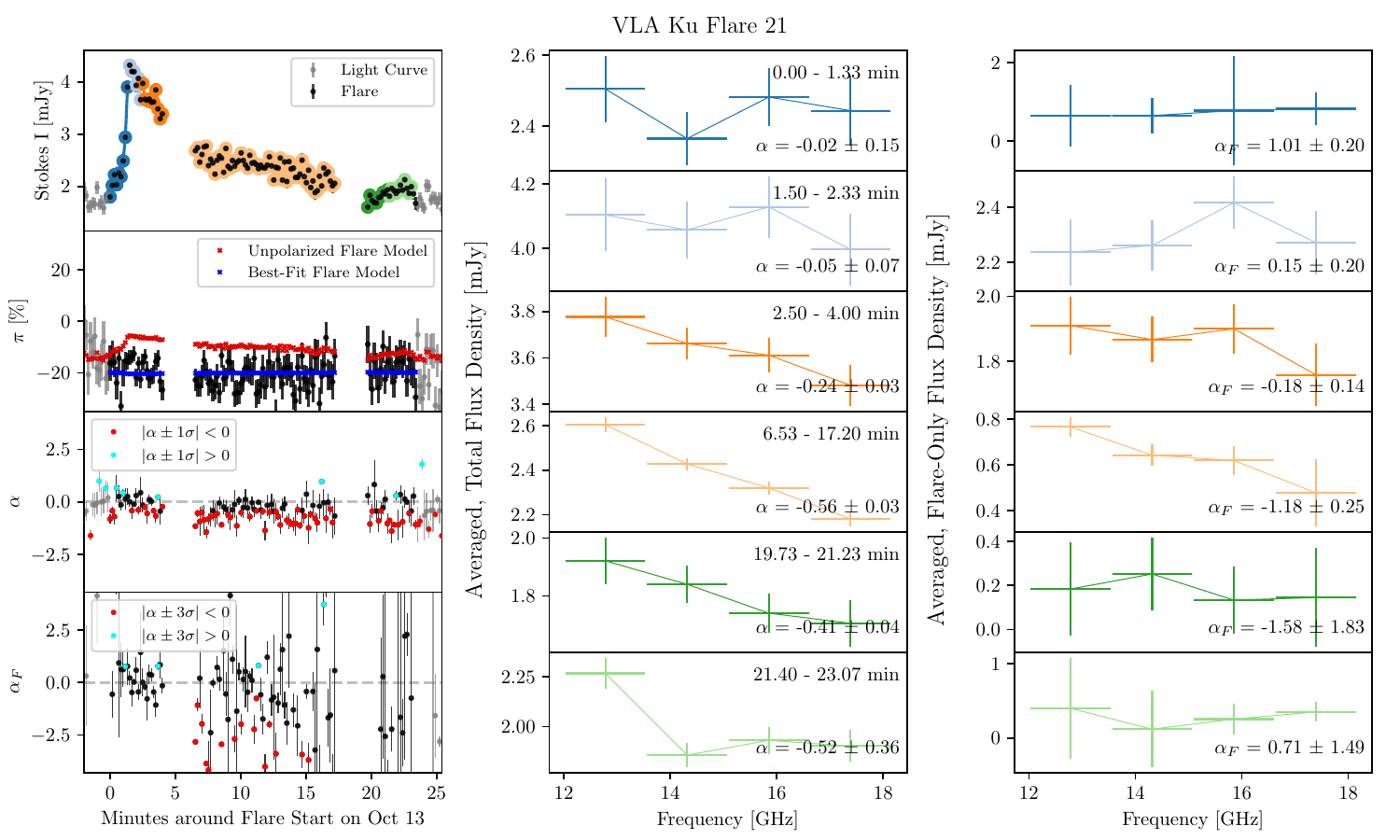}
\figsetgrpnote{Stokes I values, percentage of circularly polarized flux ($\pi$), and spectral index of the total ($\alpha$) and flare-only flux ($\alpha_F$) are shown for VLA Ku Flare 21. Descriptions are the same as in Figure \ref{fig:FigureFlare}.}
\figsetgrpend

\figsetgrpstart
\figsetgrpnum{5.2}
\figsetgrptitle{Flare 22}
\figsetplot{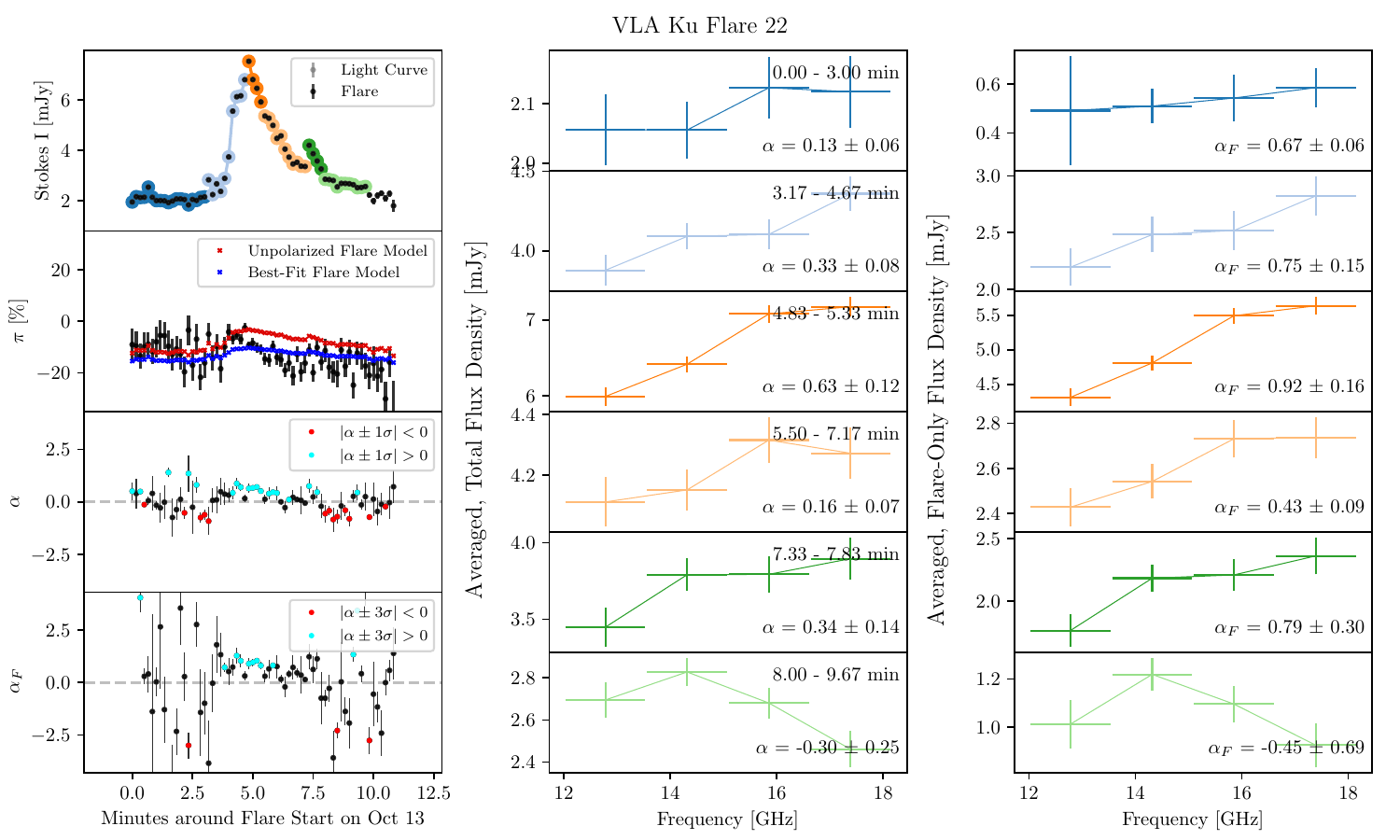}
\figsetgrpnote{Stokes I values, percentage of circularly polarized flux ($\pi$), and spectral index of the total ($\alpha$) and flare-only flux ($\alpha_F$) are shown for VLA Ku Flare 22. Descriptions are the same as in Figure \ref{fig:FigureFlare}.}
\figsetgrpend

\figsetgrpstart
\figsetgrpnum{5.3}
\figsetgrptitle{Flare 23}
\figsetplot{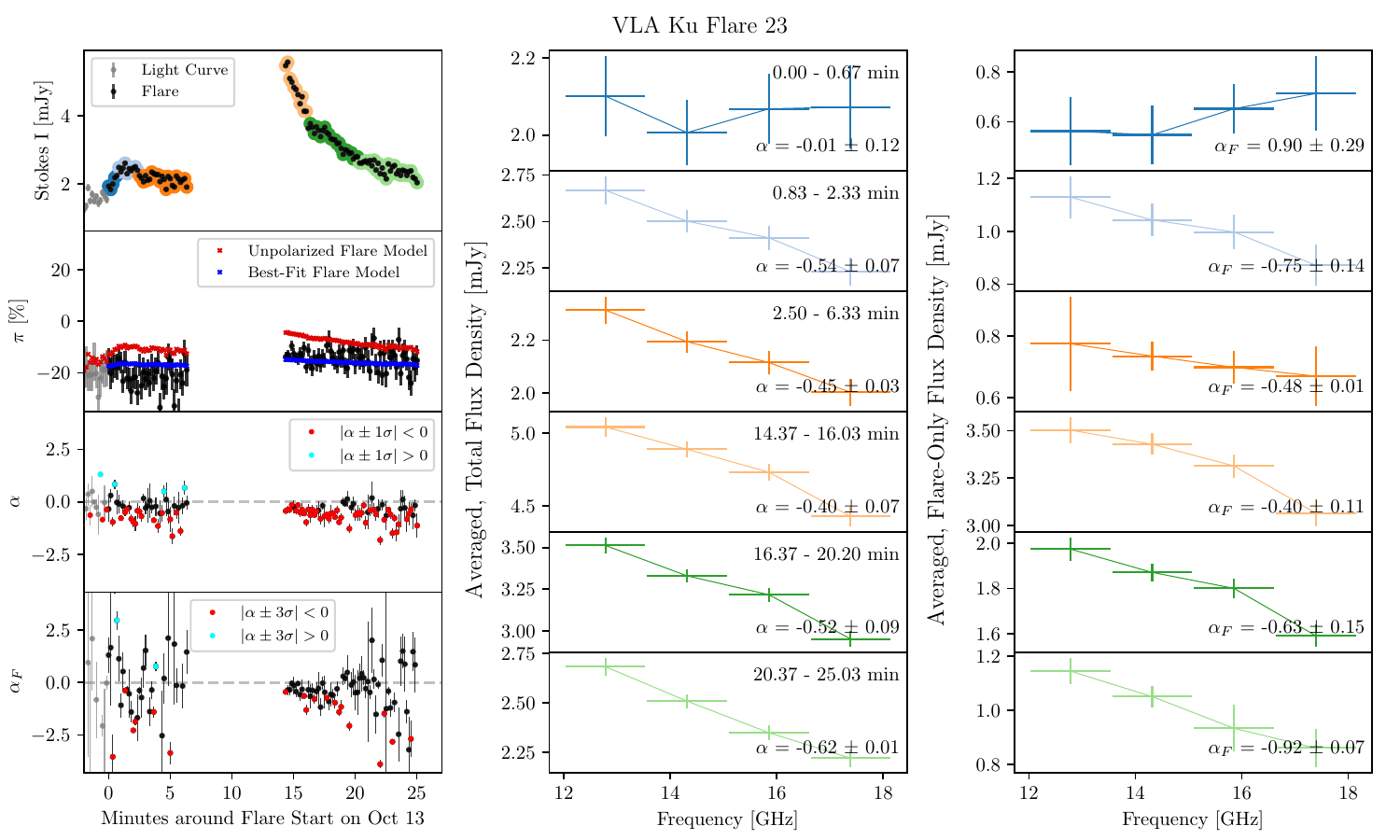}
\figsetgrpnote{Stokes I values, percentage of circularly polarized flux ($\pi$), and spectral index of the total ($\alpha$) and flare-only flux ($\alpha_F$) are shown for VLA Ku Flare 23. Descriptions are the same as in Figure \ref{fig:FigureFlare}.}
\figsetgrpend

\figsetgrpstart
\figsetgrpnum{5.4}
\figsetgrptitle{Flare 33}
\figsetplot{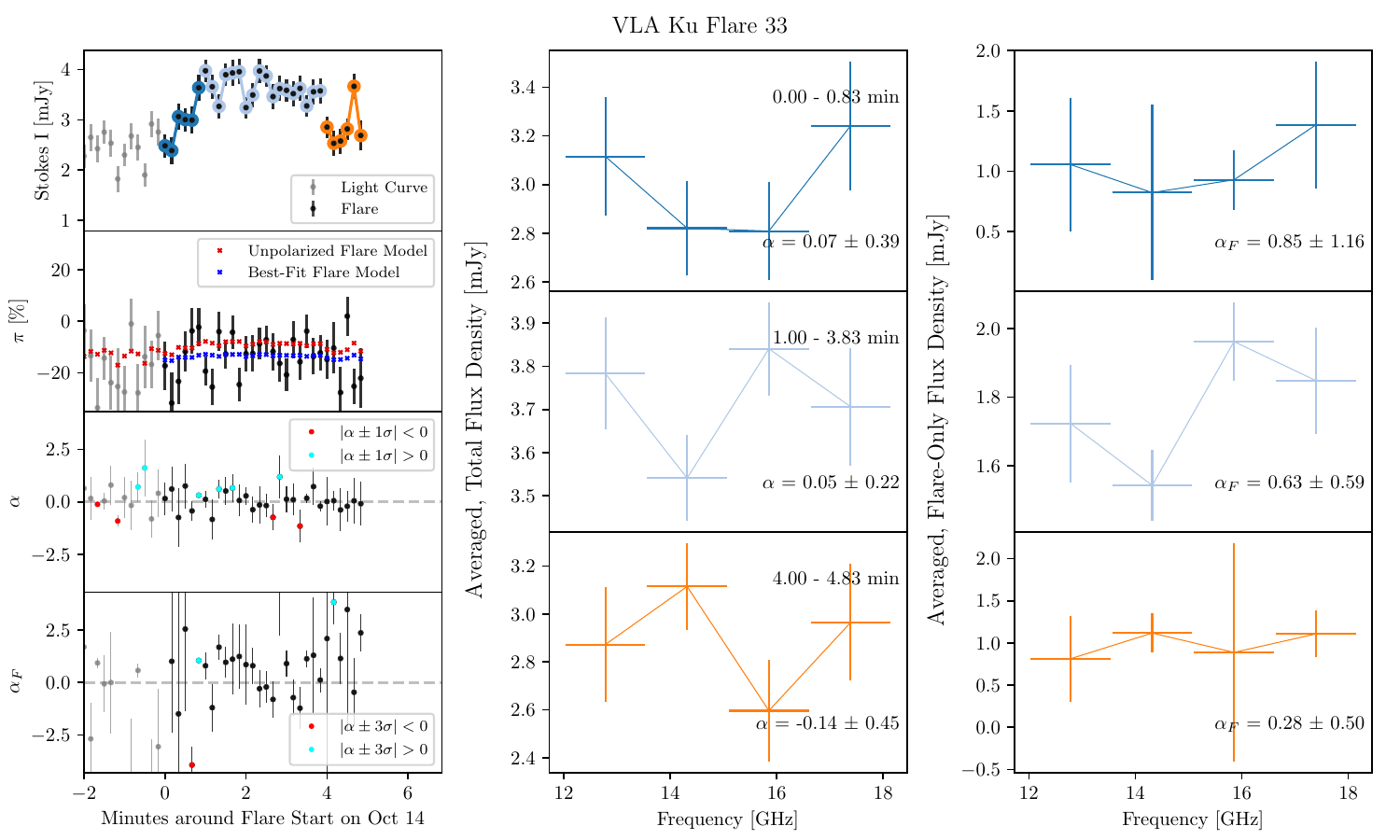}
\figsetgrpnote{Stokes I values, percentage of circularly polarized flux ($\pi$), and spectral index of the total ($\alpha$) and flare-only flux ($\alpha_F$) are shown for VLA Ku Flare 33. Descriptions are the same as in Figure \ref{fig:FigureFlare}.}
\figsetgrpend

\figsetgrpstart
\figsetgrpnum{5.5}
\figsetgrptitle{Flare 35}
\figsetplot{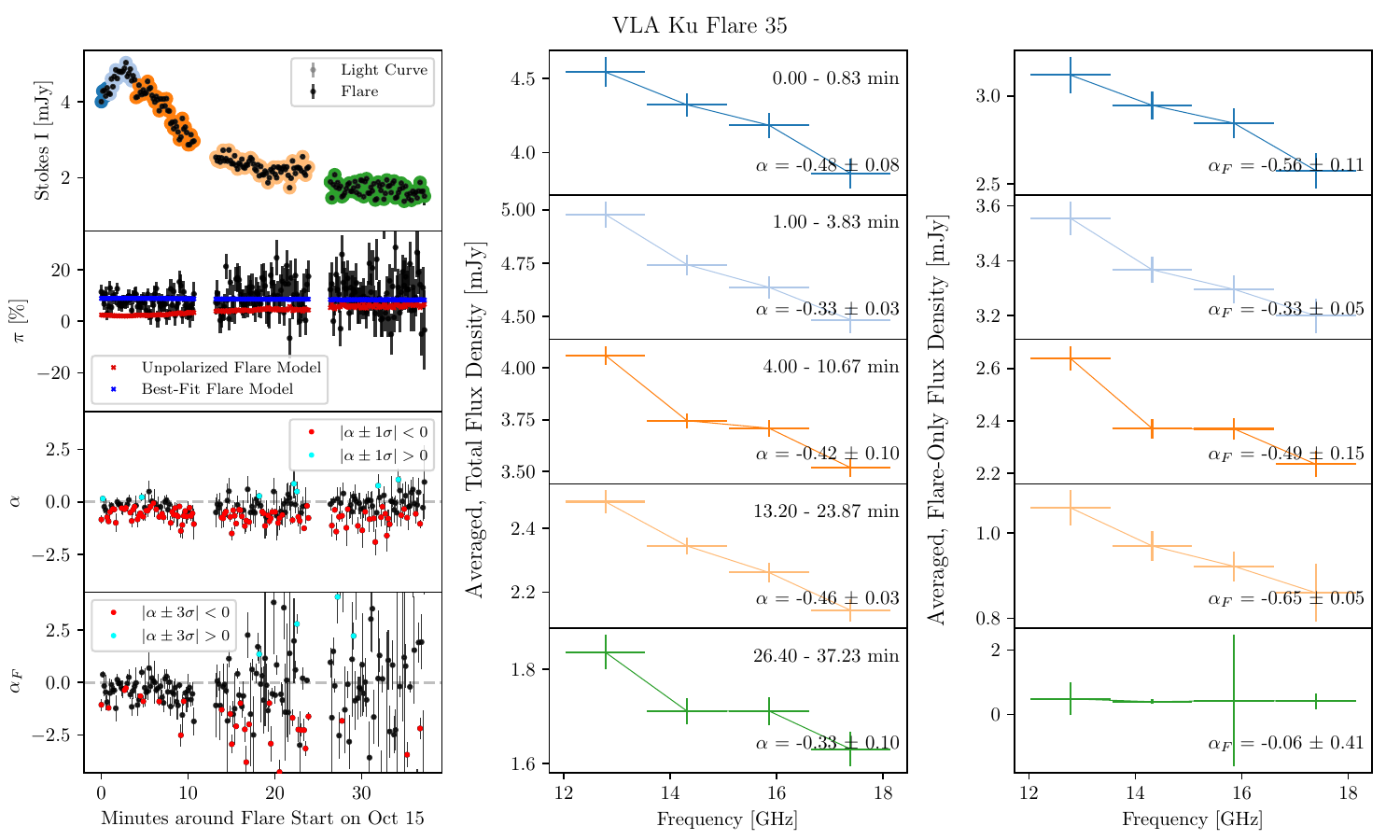}
\figsetgrpnote{Stokes I values, percentage of circularly polarized flux ($\pi$), and spectral index of the total ($\alpha$) and flare-only flux ($\alpha_F$) are shown for VLA Ku Flare 35. Descriptions are the same as in Figure \ref{fig:FigureFlare}.}
\figsetgrpend

\figsetgrpstart
\figsetgrpnum{5.6}
\figsetgrptitle{Flare 36}
\figsetplot{/FigureFlare36.pdf}
\figsetgrpnote{Stokes I values, percentage of circularly polarized flux ($\pi$), and spectral index of the total ($\alpha$) and flare-only flux ($\alpha_F$) are shown for VLA Ku Flare 36. Descriptions are the same as in Figure \ref{fig:FigureFlare}.}
\figsetgrpend

\figsetgrpstart
\figsetgrpnum{5.7}
\figsetgrptitle{Flare 74}
\figsetplot{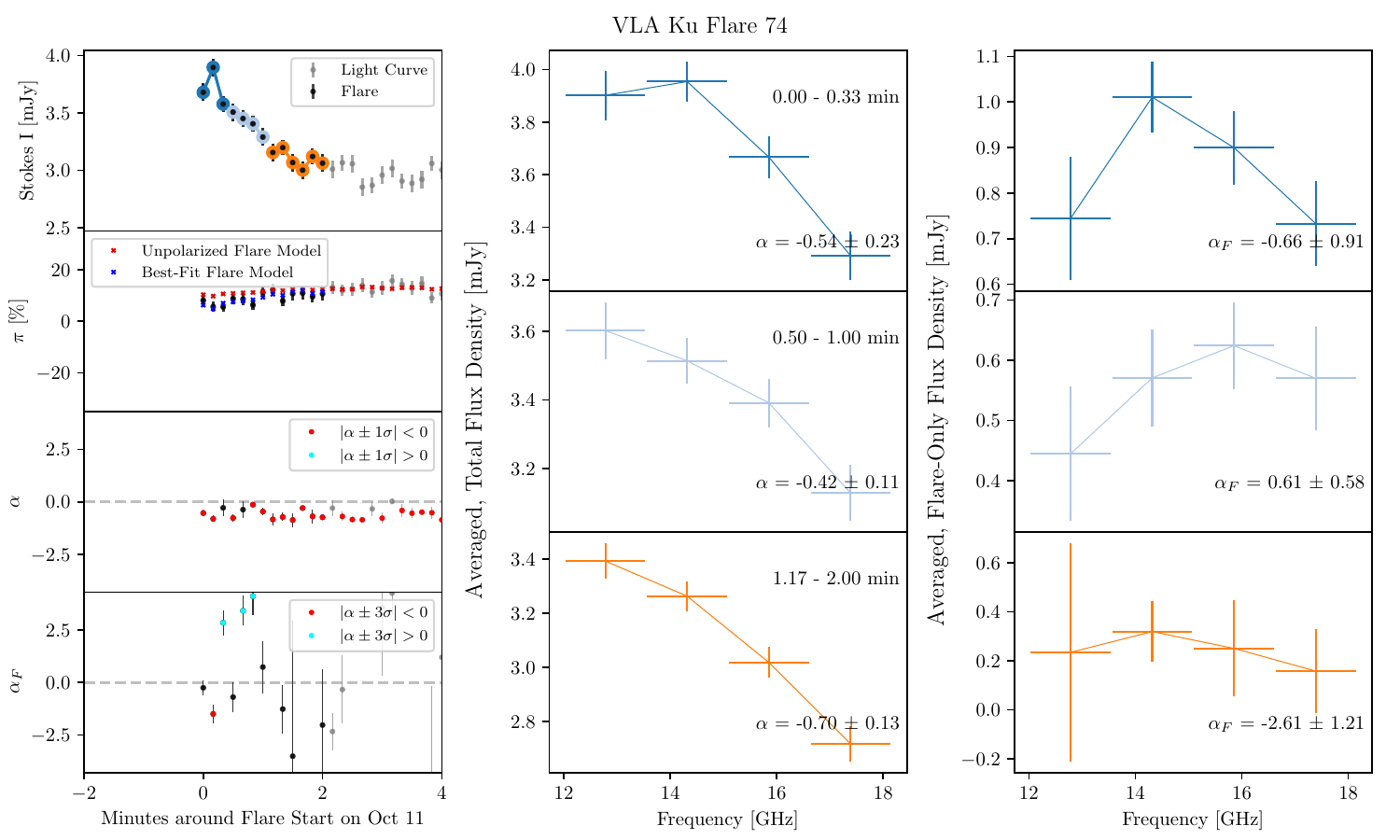}
\figsetgrpnote{Stokes I values, percentage of circularly polarized flux ($\pi$), and spectral index of the total ($\alpha$) and flare-only flux ($\alpha_F$) are shown for VLA Ku Flare 74. Descriptions are the same as in Figure \ref{fig:FigureFlare}.}
\figsetgrpend

\figsetgrpstart
\figsetgrpnum{5.8}
\figsetgrptitle{Flare 75}
\figsetplot{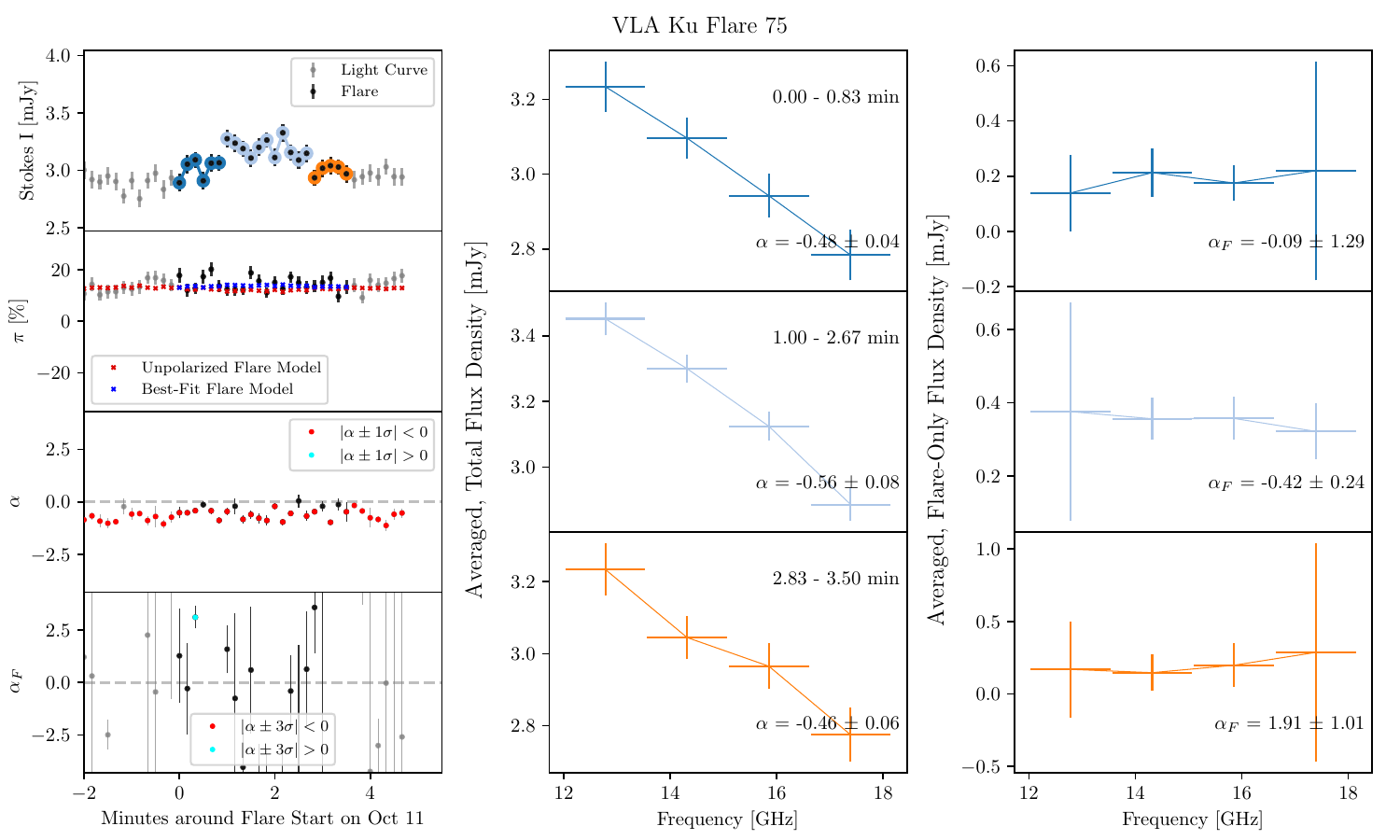}
\figsetgrpnote{Stokes I values, percentage of circularly polarized flux ($\pi$), and spectral index of the total ($\alpha$) and flare-only flux ($\alpha_F$) are shown for VLA Ku Flare 75. Descriptions are the same as in Figure \ref{fig:FigureFlare}.}
\figsetgrpend

\figsetgrpstart
\figsetgrpnum{5.9}
\figsetgrptitle{Flare 76}
\figsetplot{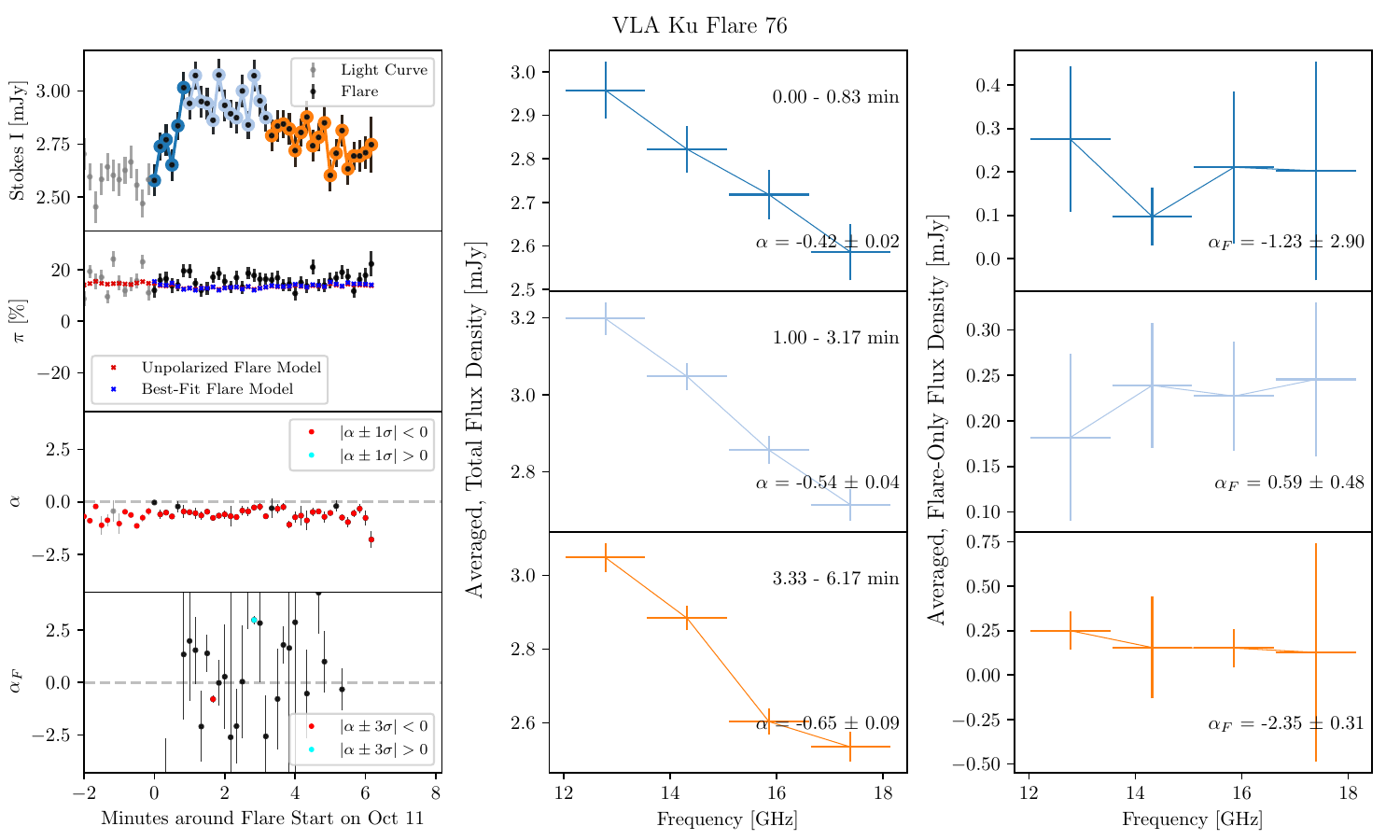}
\figsetgrpnote{Stokes I values, percentage of circularly polarized flux ($\pi$), and spectral index of the total ($\alpha$) and flare-only flux ($\alpha_F$) are shown for VLA Ku Flare 76. Descriptions are the same as in Figure \ref{fig:FigureFlare}.}
\figsetgrpend

\figsetgrpstart
\figsetgrpnum{5.10}
\figsetgrptitle{Flare 77}
\figsetplot{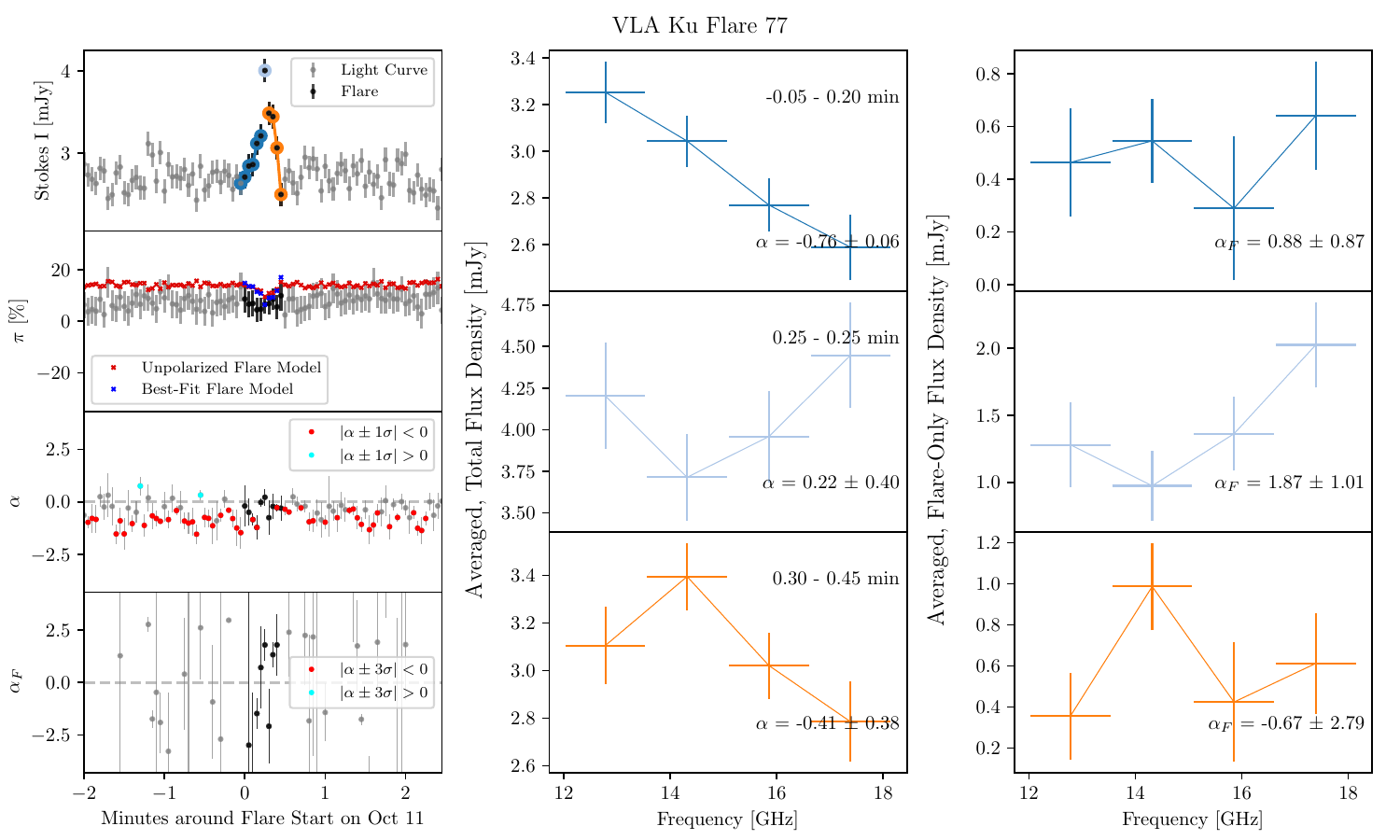}
\figsetgrpnote{Stokes I values, percentage of circularly polarized flux ($\pi$), and spectral index of the total ($\alpha$) and flare-only flux ($\alpha_F$) are shown for VLA Ku Flare 77. Descriptions are the same as in Figure \ref{fig:FigureFlare}.}
\figsetgrpend

\figsetgrpstart
\figsetgrpnum{5.11}
\figsetgrptitle{Flare 78}
\figsetplot{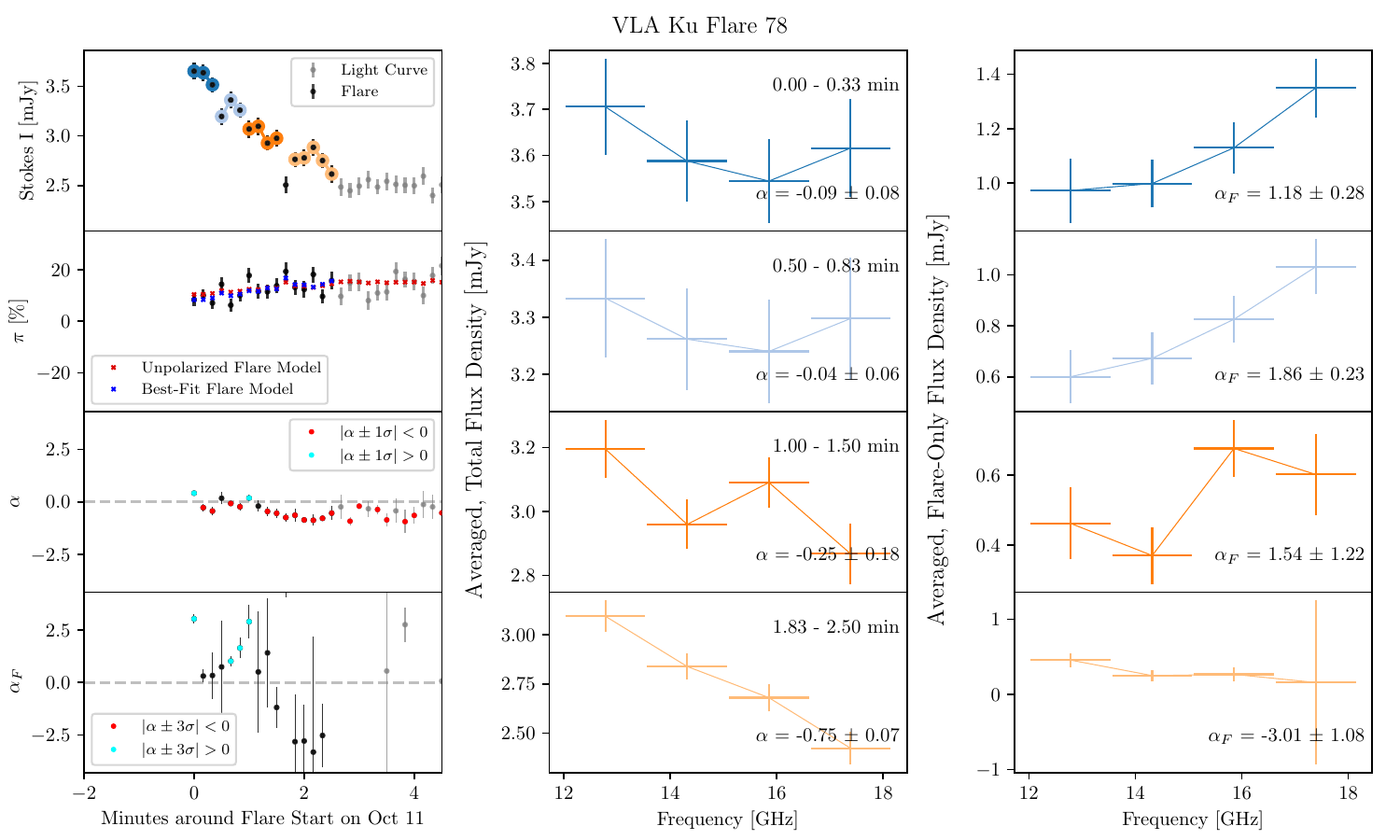}
\figsetgrpnote{Stokes I values, percentage of circularly polarized flux ($\pi$), and spectral index of the total ($\alpha$) and flare-only flux ($\alpha_F$) are shown for VLA Ku Flare 78. Descriptions are the same as in Figure \ref{fig:FigureFlare}.}
\figsetgrpend

\figsetgrpstart
\figsetgrpnum{5.12}
\figsetgrptitle{Flare 79}
\figsetplot{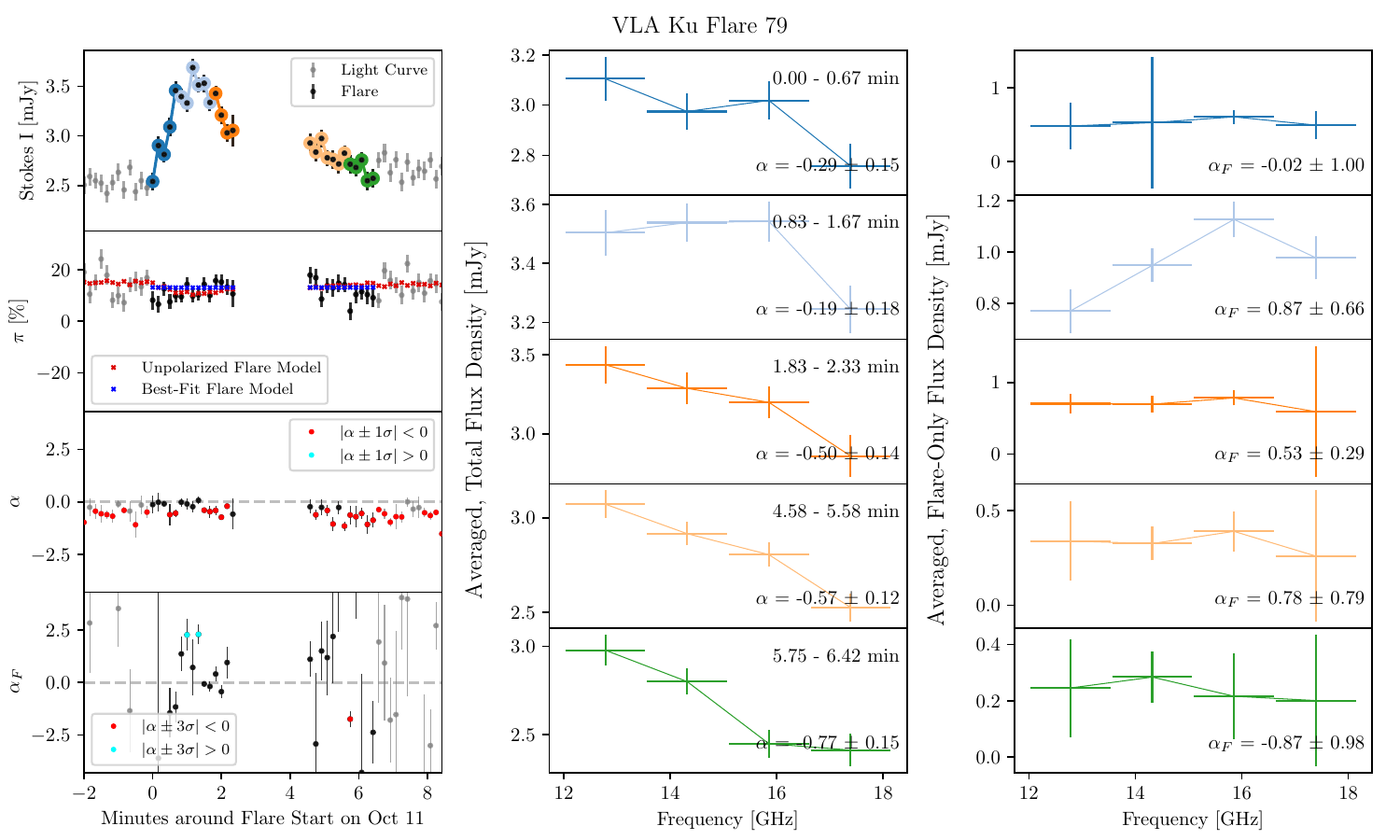}
\figsetgrpnote{Stokes I values, percentage of circularly polarized flux ($\pi$), and spectral index of the total ($\alpha$) and flare-only flux ($\alpha_F$) are shown for VLA Ku Flare 79. Descriptions are the same as in Figure \ref{fig:FigureFlare}.}
\figsetgrpend

\figsetgrpstart
\figsetgrpnum{5.13}
\figsetgrptitle{Flare 80}
\figsetplot{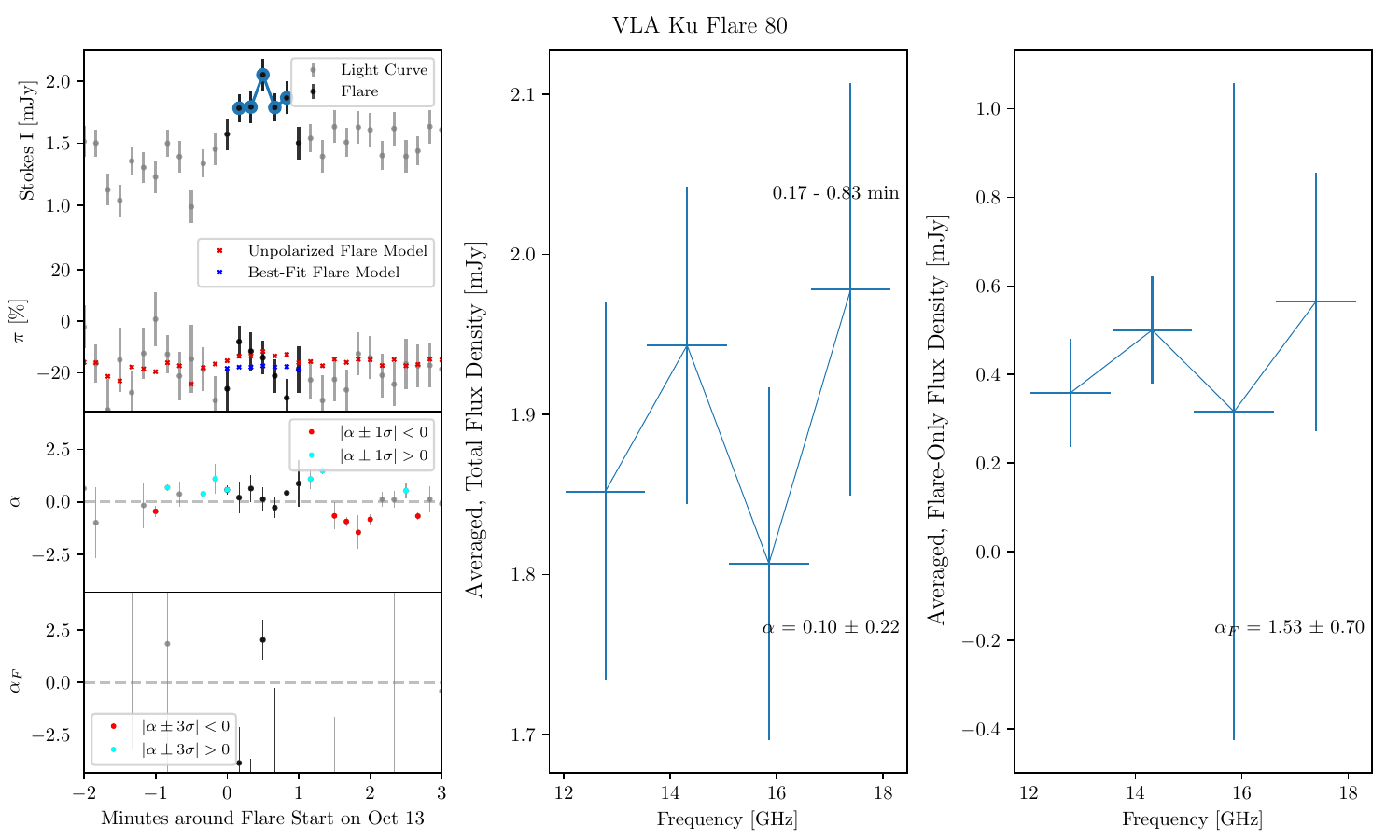}
\figsetgrpnote{Stokes I values, percentage of circularly polarized flux ($\pi$), and spectral index of the total ($\alpha$) and flare-only flux ($\alpha_F$) are shown for VLA Ku Flare 80. Descriptions are the same as in Figure \ref{fig:FigureFlare}.}
\figsetgrpend

\figsetgrpstart
\figsetgrpnum{5.14}
\figsetgrptitle{Flare 82}
\figsetplot{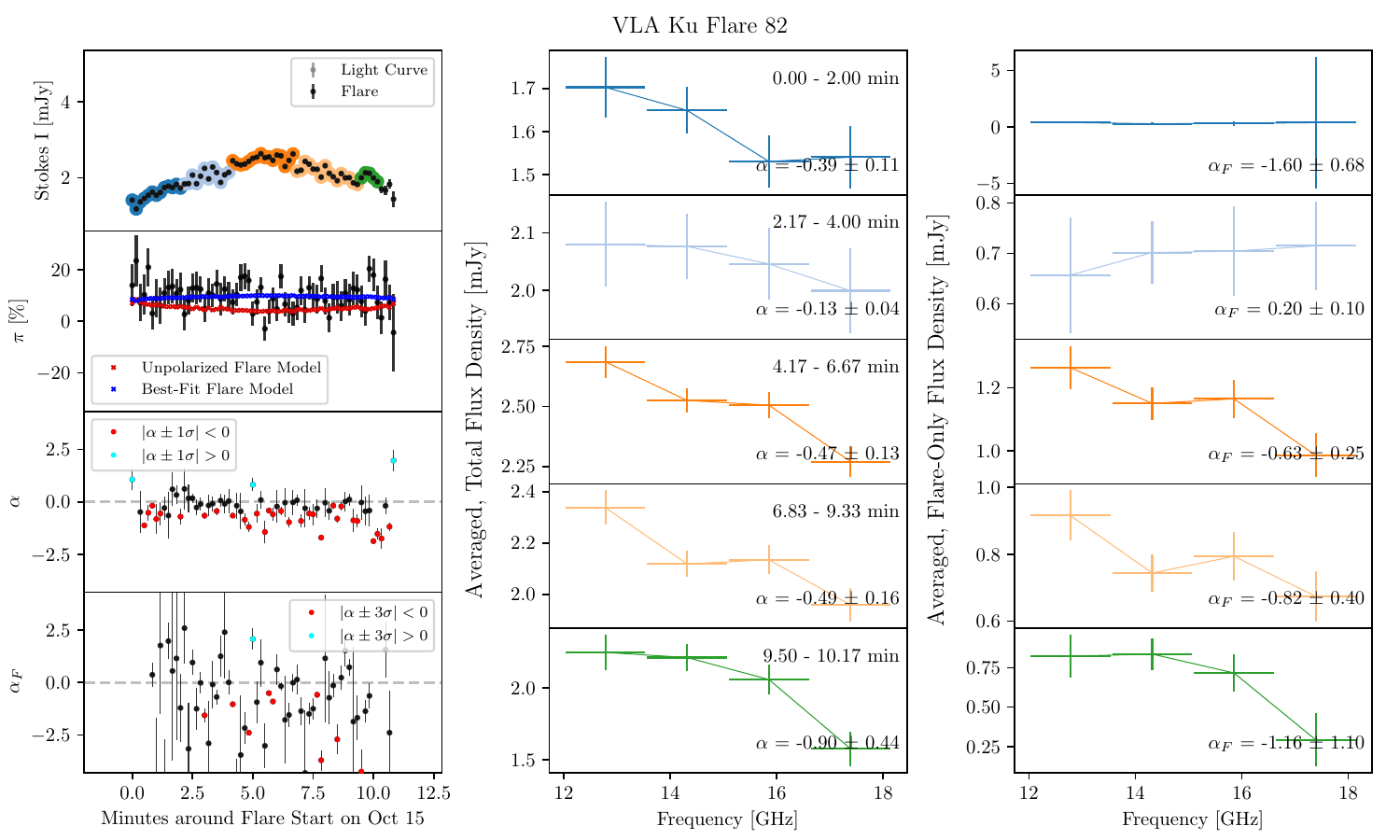}
\figsetgrpnote{Stokes I values, percentage of circularly polarized flux ($\pi$), and spectral index of the total ($\alpha$) and flare-only flux ($\alpha_F$) are shown for VLA Ku Flare 82. Descriptions are the same as in Figure \ref{fig:FigureFlare}.}
\figsetgrpend

\figsetgrpstart
\figsetgrpnum{5.15}
\figsetgrptitle{Flare 83}
\figsetplot{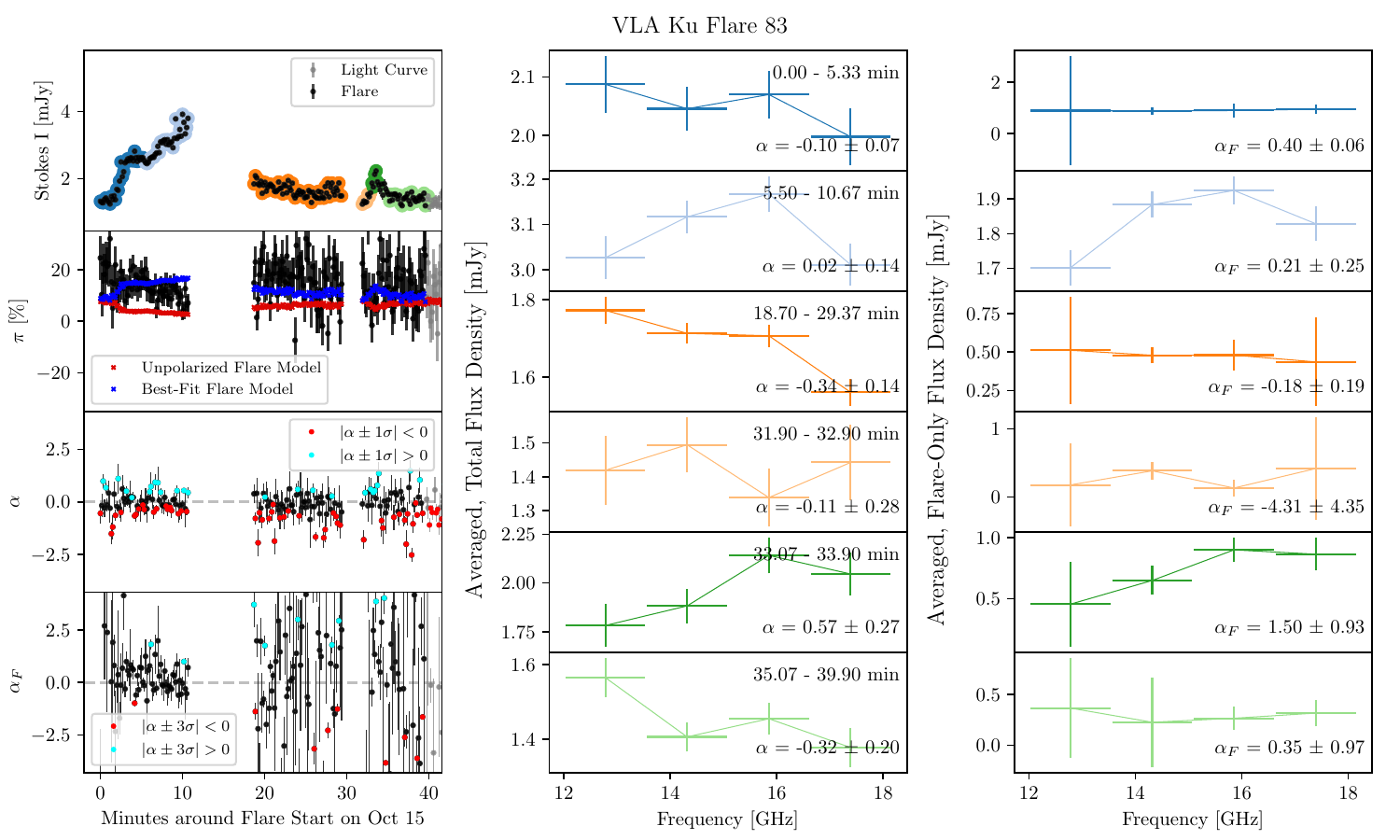}
\figsetgrpnote{Stokes I values, percentage of circularly polarized flux ($\pi$), and spectral index of the total ($\alpha$) and flare-only flux ($\alpha_F$) are shown for VLA Ku Flare 83. Descriptions are the same as in Figure \ref{fig:FigureFlare}.}
\figsetgrpend

\figsetend

\begin{figure}[!ht]
\centering
\includegraphics[width=1\linewidth]{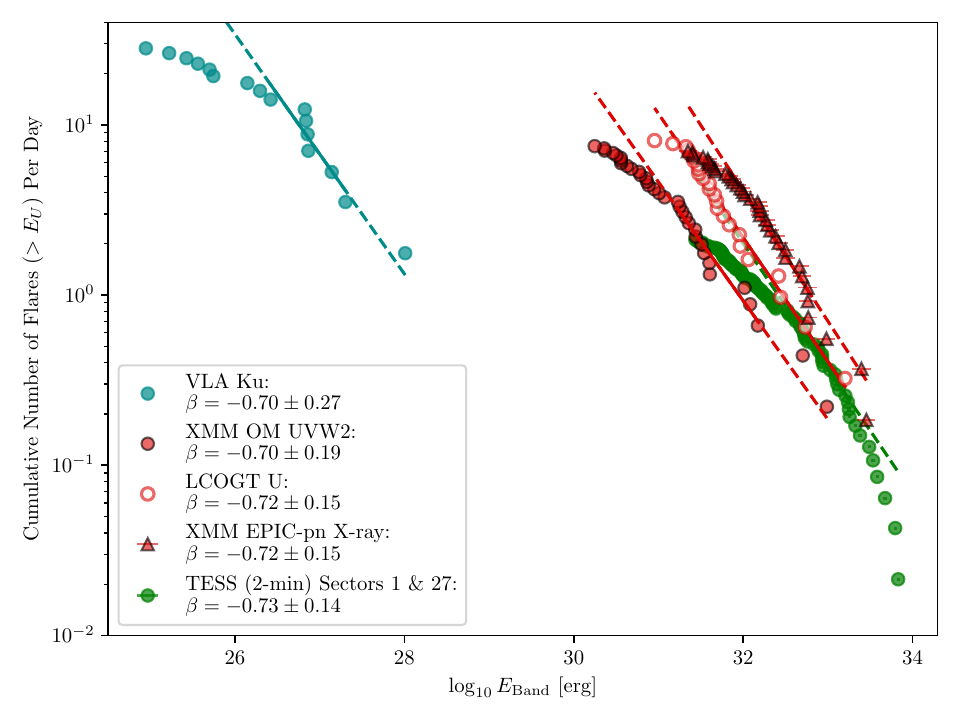}
\caption{\label{fig:FFD}
Flare frequency diagrams (FFD) of the VLA Ku-band flares compared to the FFDs of the LCOGT U, XMM EPIC-pn X-ray, and XMM OM UVW2 flares from T23. 
The flares that comprise the TESS FFD are not simultaneous with the other observations.
This FFD is shown to emphasize the similar energy range to the LCOGT U-band FFD. 
Note that the FFD slope, $\beta$, of the mid- to high-energy flares is consistent between all bands. A relatively high number of low-energy flares were observed in the VLA Ku and XMM OM UVW2 bands, and there are no clear extremely-high-energy outliers beyond the trend of these FFDs. The bandwidths used in the energy calculations are $\Delta \lambda_{\text{UVW2}} = 475$ \AA{}, $\Delta \lambda_{\text{U}} = 700$ \AA{}, and $\Delta \lambda_{\text{TESS}} = 3982$ \AA{}.
}
\end{figure}

\begin{figure}[!ht]
\centering
\includegraphics[width=1\linewidth]{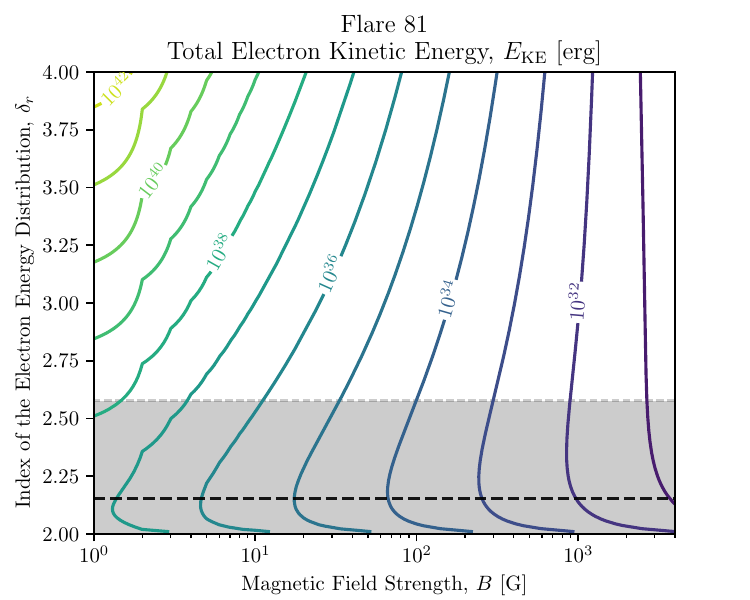}
\caption{\label{fig:Contour}
Analysis of radio flare emission in the optically thin regime for Flare 81 (see \S{\ref{sec:flare_electron_energies}}). 
Contours of the kinetic energy from the gyrosynchrotron-emitting electrons.
The contours are based on a range of reasonable magnetic field strengths and electron energy distribution indices that were calculated in the text.
The black, dashed line and gray area correspond to the index using Equation \ref{eqn:optically_thin_alpha} and 
$\alpha = -0.72 \pm 0.4$, the spectral index of Flare 81 averaged over the peak and early decay.
However, $\alpha$ varies during the flare, indicating an evolving $\delta_r$ that trends towards higher values. 
Contours for Flares 21, 23, and 35 are similar in shape, with label values reduced by an order of magnitude.
}
\end{figure}

\begin{figure}[!ht]
\centering
\includegraphics[width=1\linewidth]{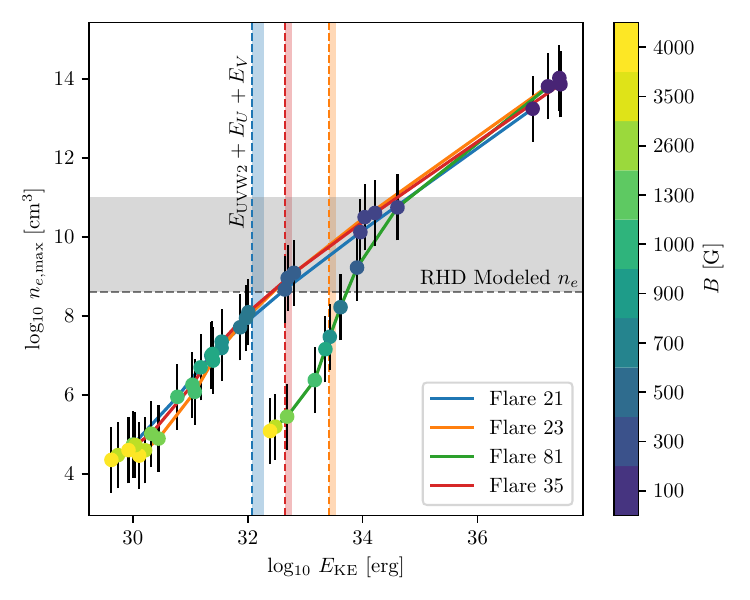}
\caption{\label{fig:EnergyEstimates}
Total electron kinetic energies of optically thin flares, based on a reasonable range of magnetic field strengths. 
Several representative nonthermal electron number densities are shown using the estimates of total flare loop volume (see text, \S{\ref{sec:flare_optically_thick}}). 
The errorbars represent the range corresponding to fractional surface area coverages of the flare footprints of 0.01 to 0.1\%{, with an additional factor for the source size uncertainty (see Figure \ref{fig:flares_geometry}). }
The vertical dashed lines indicate the total energies from the XMM OM UVW2 and LCOGT U bands for comparison with the electron kinetic energies. The vertical shaded regions of the same color indicate the energy range when including the LCOGT or SMARTS V-band (see T23). Note that the solid and dashed line colors correspond to Flare IDs. The gray dashed line corresponds to the number density of electrons employed in recent radiative-hydrodynamic (RHD) simulations of M-dwarf flares \citep{Kowalski2025}.
For {models with} smaller low-energy cutoffs, the number density {can approach $10^{11}$} cm$^{-3}$ which is shown {by} the gray region. Extended discussion is presented in \S{\ref{sec:disc_kinetic_energies}}.
}
\end{figure}

\begin{figure}[!ht]
\centering
\includegraphics[width=0.5\linewidth]{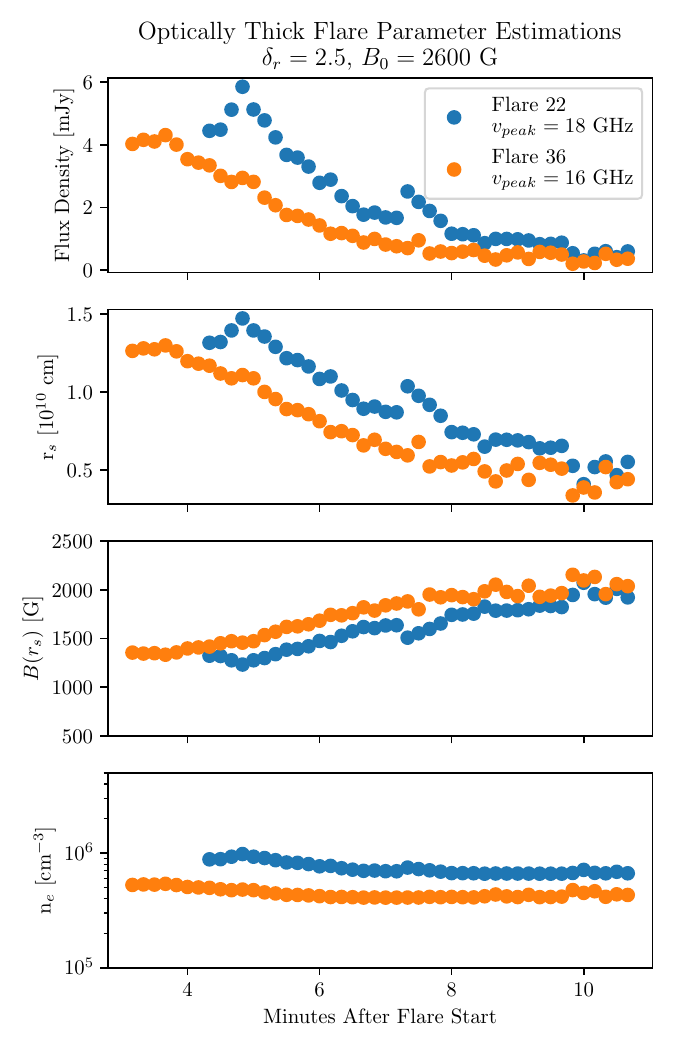}
\caption{\label{fig:flares_optically_thick}
Analysis of radio flare emission in the optically thick regime (see \S{\ref{sec:flare_optically_thick}}). Estimations of the source size scale ($r_s$), minimum magnetic field strength at the top of the source ($B(r_s)$), and the electron density ($n_e$) are shown for Flares 22 and 36 after the rise phase. A photospheric value of $B_0 = 2600$ G is assumed, and a power-law index of $\delta_r = 2.5$ is obtained from the analysis of the optically thin flares. The peak frequency ($\nu_{\text{peak}}$) is constrained by the observed peak in the Ku band (see Figure Set 1). Due to the flare size, high values for $B(r_s)$ and $n_e$ are necessary to explain the flux densities observed. 
}
\end{figure}

\begin{figure}[!ht]
\centering
\includegraphics[width=1.0\linewidth]{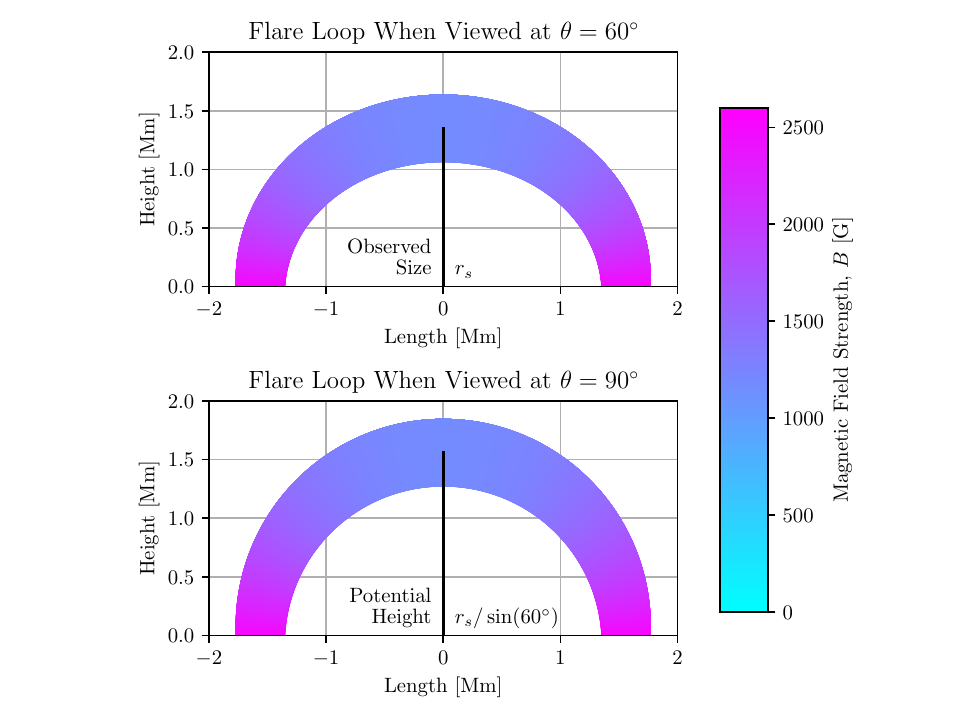}
\caption{\label{fig:flares_geometry}
{Illustration of assumed geometry between optically thick and thin flares at two angles. Flares are assumed to come from plasma following a semi-circular magnetic loop, such that height would equal the radius of the loop. The optically thick layer, which is probed for the source size scale ($r_s$), is assumed to be centered within the apex of the loop and roughly equal to the loop height. Given a line-of-sight angle of $60$\degr{}, the total volume may be underestimated by about 13\%. This is propagated through the calculations and included in the uncertainties in Figure \ref{fig:EnergyEstimates}.
}
}
\end{figure}

\begin{figure}[!ht]
\centering
\includegraphics[width=1\linewidth]{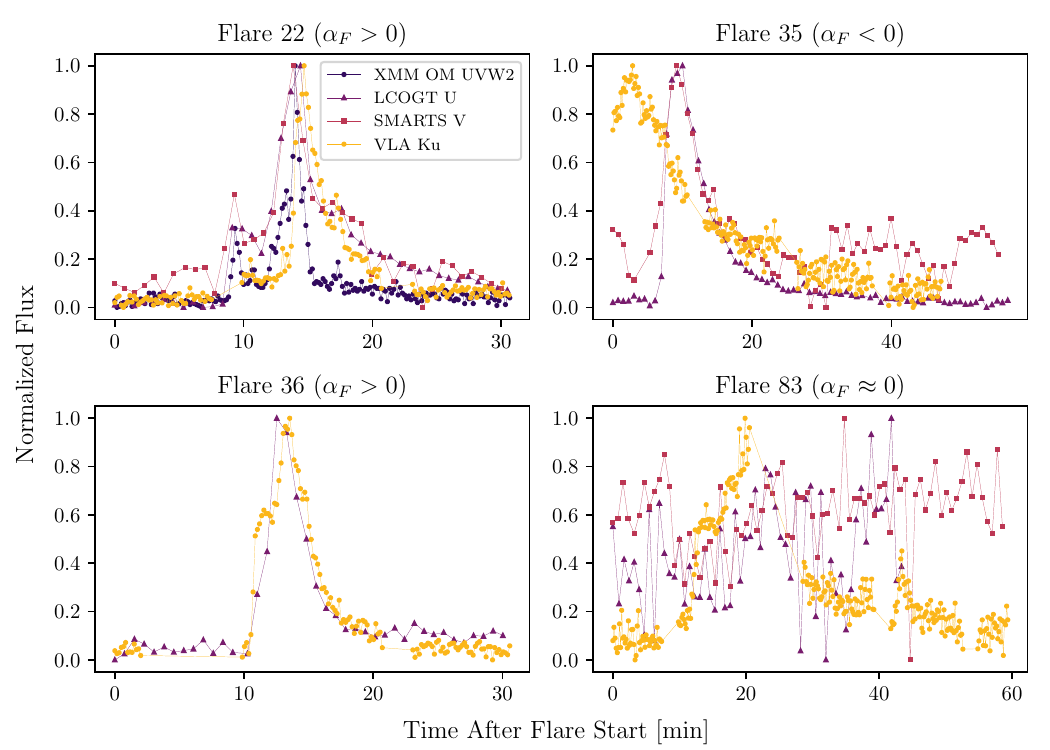}
\caption{\label{fig:flares_zoomed_in}
Comparisons of several radio flares to their multi-wavelength counterparts in T23. 
Flux density is normalized between 0 and 1 for the times shown in these intervals.
Flares 22, 35, and 36 demonstrate the close temporal correlation between the radio and other impulsive flaring emissions, and Flare 83 demonstrates a radio flare with no significant optical counterparts.
}
\end{figure}

\clearpage

\acknowledgements
I.I.T., A.F.K., and Y.N.\ acknowledge funding support through NASA ADAP award  Number 80NSSC21K0632 and NASA XMM-Newton Guest Observer AO-17 Award 80NSSC19K0665.
I.I.T.\ acknowledges support from the NSF Graduate Research Fellowship Program (GRFP). 
Y.N.\ acknowledges funding from NASA TESS Cycle 6 80NSSC24K0493, NASA NICER Cycle 6 80NSSC24K1194, and HST GO 17464.

We thank Drs.\ Rodrigo H.\ Hinojosa, Wei-Chun Jao, Jamie R.\ Lomax, James E.\ Neff, Leonardo A.\ Paredes, and Jack Soutter for their contributions during early parts of the 2018 AU Mic Campaign. {We also thank an anonymous reviewer for helpful comments that led to the improvement of this manuscript.} I.I.T.\ thanks Dr.\ Jackie Villadsen and Dr.\ Laura D.\ Vega for useful discussion on M-dwarf radio data and Imogen G.\ Cresswell for early contributions to the development of EffFD. 

The National Radio Astronomy Observatory is a facility of the National Science Foundation operated under cooperative agreement by Associated Universities, Inc. 
The Australia Telescope Compact Array is part of the Australia Telescope National Facility (grid.421683.a) which is funded by the Australian Government for operation as a National Facility managed by CSIRO. We acknowledge the Gomeroi people as the traditional owners of the Observatory site. 
This work is based on observations obtained with XMM-Newton, an ESA science mission with instruments and contributions directly funded by ESA Member States and NASA.
This work makes use of observations from the Las Cumbres Observatory global telescope network.
This research has used data from the SMARTS 1.5m and 0.9m telescopes, which are operated as part of the SMARTS Consortium.
This research made use of Lightkurve, a Python package for Kepler and TESS data analysis (Lightkurve Collaboration, 2018).

\appendix
\section{VLA Data Calibration} \label{sec:appendix_vla_calibration}

Additional VLA data handling and calibrations are described here for reference.
The original observation logs are available from the NRAO log repository\footnote{\url{http://www.vla.nrao.edu/cgi-bin/oplogs.cgi}} (code 18B-193).
The on-source observation times used after the data reduction (for both the VLA and ATCA) are available in Table \ref{tbl:obs_logs}.
Notably, data from antennas \texttt{ea22} (Oct 11), \texttt{ea13} and \texttt{ea1} (Oct 15), and \texttt{ea17} (all days) were corrupted, resulting in data loss. 
Antenna \texttt{ea10} was used in different observations during Oct 14 and 15.
The weather was relatively clear on Oct 11, but cloudy or overcast on other days. 
The relative humidity was somewhat high during these observations, with a steady value of 40\% during Oct 11, but starting from 60\% and settling around 80--85\% after two hours during the other nights. The \texttt{plotweather} function in \texttt{CASA} reveals the opacity correction to be consistent with the standard seasonal weather model for Oct 13--15 and slightly lower for Oct 11. There were no times of extreme, temporary relative humidity increases.

VLA data are calibrated following the CASA VLA Continuum Tutorial from NRAO\footnote{\url{https://casaguides.nrao.edu/index.php?title=VLA_Continuum_Tutorial_3C391-CASA6.4.1}}.
RFI is manually searched for and removed from all data before starting calibration.
This is done in a 2-pass cycle.
First, values above 500 times the standard range are clipped.
Then, spikes only affecting small channel ranges above 100 times the average value are removed.
The standard flags of clipping zeros, quacking, and shadowing are applied.
Erratic data are removed from channel 20 of spectral window 47, channels 110-111 of spectral window 30, and baselines \texttt{ea10\&ea12} and \texttt{ea12\&ea23} for Oct 11 and 14.
While this greatly improves the diagnostic images for these spectral windows, the effects on the final light curve are negligible, as the affected data volume is low. 
Notably, we add gain and opacity calibrations and the solution intervals for the gain and phase calibrators are changed to 60s to perform better calibrations given any atmospheric variations. Both are these are to mitigate weather effects as much as possible since the Ku band is sensitive to weather conditions. Polarization from flares is expected to dominate the Stokes $V$ parameter during events, so extra calibrations for polarization are not required.

The reported flux density from the NRAO for J2040-2507 is 0.40 Jy, with up to 10\% expected error due to the quality of the calibrator. 
The reported flux density values per day in chronological order are 439.424 $\pm$ 0.548816, 488.691 $\pm$ 0.768659, 573.353 $\pm$ 1.01744, 506.878  $\pm$ 0.714263 mJy.
Thus, all days are within expected error despite the variability.
The reported rms values are slightly different for Oct 11 and Oct 14, which are likely due to better and worse weather conditions, respectively.
Note that these do not reflect the changes in average quiescence per day.

During the analysis, \texttt{uvmodelfit} is used with an initial guess of fixed position $x,y=0$ and $I=1.5$ mJy (based on initial tests) with 5 iterations for all days. Noise is not decreased when testing varied positions and 10/15 iterations. For reference, \texttt{CASA} 6.6.4 does not save the errors in output files, so values are taken from the printed output and matched to the timebinned data. These errors are consistent with time-integrated estimations using \texttt{tclean} and \texttt{imfit} over quiescent and flare FWHM times, with slight variability based on how much data is used in \texttt{uvmodelfit} (e.g.\ 3-sec or 10-sec binning).

\begin{deluxetable}{ccccc}[!ht]
\tabletypesize{\scriptsize}
\tablewidth{0pt}
\tablecaption{Observation Logs \label{tbl:obs_logs}}
\tablehead{
\colhead{Instrument} & \colhead{Observation Window} & \colhead{Start} & \colhead{End} & \colhead{Duration} \\
 & & (UTC) & (UTC) & [hrs]}
\startdata 
VLA & 1 & 2018-10-11T01:04:18.0 & 2018-10-11T01:15:27.0 & 0.19 \\
VLA & 2 & 2018-10-11T01:17:30.0 & 2018-10-11T01:28:42.0 & 0.19 \\
VLA & 3 & 2018-10-11T01:30:45.0 & 2018-10-11T01:41:54.0 & 0.19 \\
\nodata & \nodata & \nodata & \nodata & \nodata \\
\enddata
\tablecomments{Observation windows for the VLA and ATCA data. Windows are separated by gaps greater than 2 minutes in the VLA data and 5 minutes in the ATCA data. This is an example log, with the full table available online in a machine-readable format.}
\end{deluxetable}

\section{TESS Data Analysis of AU Mic}
\label{sec:appendix_tess}
To quickly analyze TESS flares, we developed the Efficient Flare Distributions (EffFD) Python code\footnote{\url{https://github.com/isatri/EffFD}}. Given an input list of stars, this program automatically pulls all available TESS data using \texttt{lightkurve} \citep{Lightkurve2018}, de-trends the \texttt{PDCSAP-FLUX} light curve, returns a list of identified flares, and builds an FFD based on the equivalent durations (ED) in the TESS band. The goal of this program was to produce FFDs using quick and data-centric methods. 

We employ an iterative, two-step process to de-trend the AU Mic Sector 1 and 27 light curves (2018 July 25 to August 22 and 2020 July 5 to July 30, respectively)\footnote{All the TESS data used in this paper can be found in MAST: \dataset[10.17909/rdd5-r118]{http://dx.doi.org/10.17909/rdd5-r118}.}. The 2-minute cadence TESS data is used, as this allows for less low-level, flare-like noise ambiguity. First, we apply a Savitzky–Golay filter over a 1-hour window to flatten out the light curve. We also remove time ranges with many data gaps, which can skew the underlying quiescent line from this filter. Second, any times where the flux above $1.8 \sigma$ or below $-1 \sigma$ are flagged. The original light curve is reloaded without these times, and this process is repeated 5 times. In testing, many M-dwarfs responded to a 1.7-hour window and bottom cutoff of $-0.8 \sigma$. However, due to AU Mic's rotational period and planetary transits, a smaller flattening window and less stringent cutoff allow for better quiescent fitting. 

Candidate flares are chosen by analyzing previously removed points that rise above $3\sigma$ of the de-trended light curve. If there are 3 consecutive times that meet this criteria, the flare is automatically added and expanded in time until the flare falls to the quiescent. For shorter increases, the time-expansion is applied first; if the candidate flare is at least 8 minutes long, the flux increase is marked as a flare. These flares are then checked by eye. This method returns 49 flares in Sector 1 and 50 flares in Sector 27 over a total of 46.8 days of observations.

\begin{figure}[!ht]
\centering
\includegraphics[width=1\linewidth]{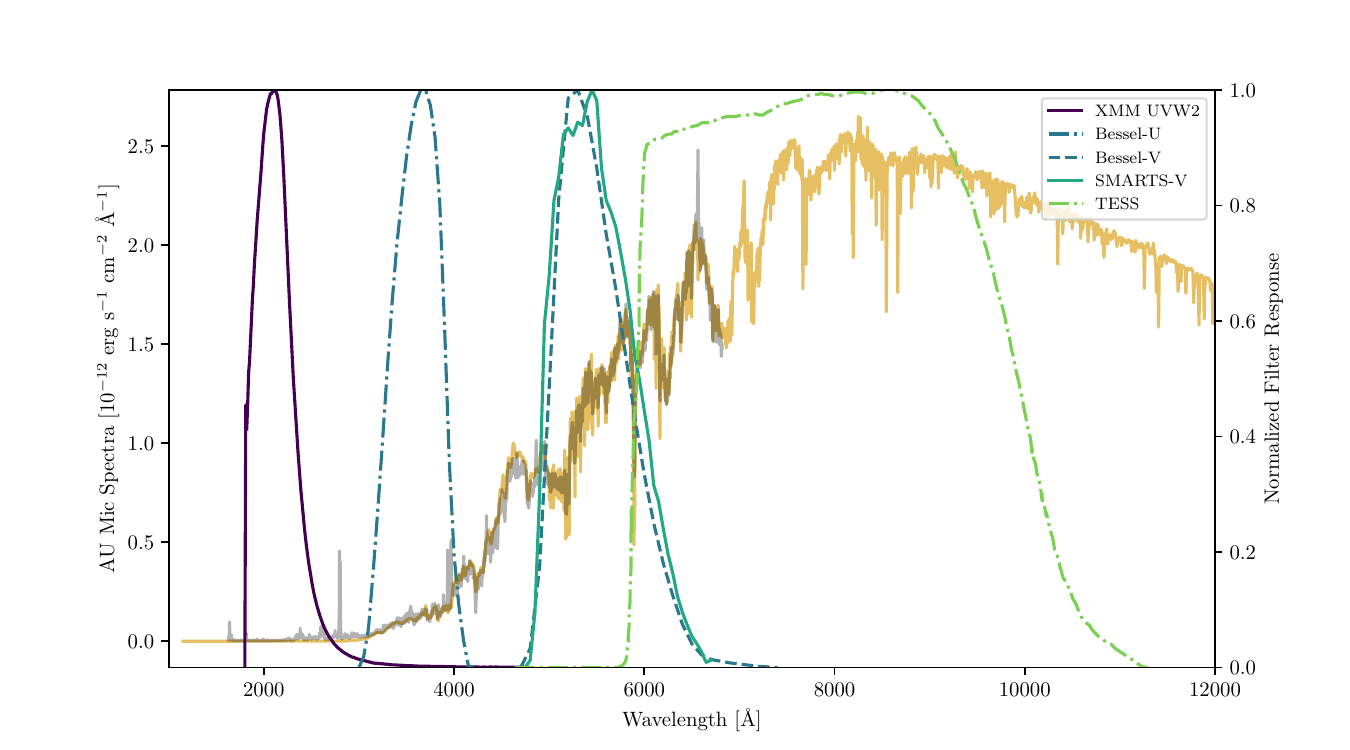}
\caption{\label{fig:filters}
Effective area or filter transmission curves (photon weighting) for the broadband filters used in this study plotted against the quiescent HST/FOS spectrum of AU Mic (gray line). The dark yellow line shows an M1-dwarf spectrum template scaled to the V-band wavelengths of the AU Mic spectrum.
All filter responses are normalized from 0 to 1.
}
\end{figure}

Energies in the TESS band are calculated using $E_{\text{TESS}} = \text{ED} \times L_{q\text{,TESS}}$, where $L_q$ is the quiescent luminosity (see \S{2.3} of T23 for a more in-depth review), which results in the TESS FFD shown in Figure \ref{fig:FFD}. The HST/COS quiescent spectrum of AU Mic \citep{Augereau2006, Tristan2023} is used to scale a dM1e optical template \citep{Bochanski2007} to the V-band wavelengths (5,000 to 6,500 \AA{}), which is then convolved with the TESS bandpass to estimate the quiescent flux in the TESS band (Figure \ref{fig:filters}). This is multiplied by $4 \pi d^2$ and the TESS bandpass FWHM (3982 \AA{}) to estimate $L_{q\text{,TESS}} = (1.1 \pm 0.05) \times 10^{32}$ erg. This quiescent luminosity (along with the flares per sector) is very similar to another estimated in a previously published TESS flaring analysis of AU Mic \citep{Gilbert2022}. However, while the extremities of the flare frequencies and energies appear consistent between methods, there is a discrepancy between energies which widens towards the lower-mid range. Differences are about half an order of magnitude. This is likely due to the difference in calculating energies (i.e.\ modeling flare shapes versus our data-only approach). These uncertainties will affect the slope of the FFD, so we use only the upper-mid to lower-high energies in calculations here. This range returns a slope similar to the other multi-wavelength observations.

\clearpage
\bibliography{tristan2024}

\end{document}